\documentclass[11pt,a4paper]{article}
\pdfoutput=1
\usepackage{jheppub}
\usepackage{bbm}
\usepackage{tikz}
\usetikzlibrary{trees}
\usepackage{tikz-qtree}
\usepackage{graphicx,subcaption}
\usepackage{mathtools}
\usepackage{bm}
\usepackage{enumitem}
\usepackage{verbatim}
\usepackage{amssymb}
\usepackage{slashed}
\usepackage{float}
\usepackage[scr = dutchcal]{mathalpha}
\usepackage{lipsum}
\usepackage{multirow}
\usepackage{hyperref}

\usepackage[normalem]{ulem}
\usepackage{orcidlink}

\newcommand{\bfmtb}{\bfmtb}
\renewcommand{\bfmtb}{\lambda}

\DeclareMathOperator{\Tr}{Tr}

\DeclareMathAlphabet{\mathbfsf}{OT1}{cmss}{bx}{n}

\newcommand{\wgt}{\text{wt}}

\newcommand{\CA}{\mathcal{A}}
\newcommand{\CB}{\mathcal{B}}

\newcommand{\CD}{\mathcal{D}}

\newcommand{\CH}{\mathcal{H}}

\newcommand{\CO}{\mathcal{O}}
\newcommand{\CP}{\mathcal{P}}

\newcommand{\CW}{\mathcal{W}}

\newcommand{\SCO}{\mathscr{O}}

\newcommand{\del}{\partial}
\newcommand{\bb}[1]{\mathbb{#1}}

\newcommand{\Zint}{\mathbb{Z}}

\newcommand{\Z}{\mathbb{Z}}

\newcommand{\Q}{\mathbf{Q}}

\renewcommand{\=}{\;= \;}
\newcommand{\+}{\;+ \;}

\newcommand\be{\begin{equation}}
\newcommand\ee{\end{equation}}
\newcommand\beq{\begin{equation}}
\newcommand\eeq{\end{equation}}
\newcommand\bea{\begin{eqnarray}}
\newcommand\eea{\end{eqnarray}}

\renewcommand{\a}{\alpha}
\renewcommand{\b}{\beta}

\renewcommand{\t}{\tau}

\newcommand{\p}{\partial}

\newcommand{\Ione}{{L_0 - c/24}}

\newcommand{\rd}{{\rm d}}

\newcommand{\identity}{\mathbbm{1}}

\renewcommand{\tilde}{\widetilde}

\newcommand{\htilde}{{\tilde h}}
\newcommand{\qhat}{{\hat q}}

\newcommand{\vev}[1]{\langle {#1} \rangle } 
\newcommand{\ket}[1]{|#1\rangle}
\newcommand{\vac}{{\ket{0}}}

\newcommand{\SD}[1]{{D^{(#1)}}}
\title{Modular Properties of {\boldmath $\CW_3$}  Generalised Gibbs Ensembles}

\author[1]{ Max Downing\orcidlink{0000-0002-8978-1151}}
\author[2]{, Faisal Karimi\orcidlink{0009-0003-5591-9911}}
\author[3,4,5]{, Tanmoy Sengupta\orcidlink{0000-0003-0220-8705}}
\author[4,5]{, Adarsh Sudhakar\orcidlink{0009-0007-4817-8659}}
\author[2]{, G\'erard M.T. Watts\orcidlink{0000-0002-9066-2838}}

\affiliation[1]{Laboratoire de Physique de l'\'Ecole Normale Sup\'erieure, \\
ENS, Universit\'e PSL, CNRS, Sorbonne Universit\'e,\\
Universit\'e Paris Cit\'e, F-75005 Paris, France\\}

\affiliation[2]{Department of Mathematics,\\
King's College London,\\
Strand, London, WC2R 2LS, United Kingdom\\}

\affiliation[3]{Chennai Mathematical Institute,\\ H1, SIPCOT IT Park, Siruseri,\\ Kelambakkam 603103, India\\}

\affiliation[4]{The Institute of Mathematical Sciences,\\
IV Cross Road, C.I.T. Campus,\\
Taramani, Chennai, India 600113\\}

\affiliation[5]{Homi Bhabha National Institute,\\
Training School Complex, Anushakti Nagar,\\
Mumbai, India 400094\\}

\emailAdd{max.downing@phys.ens.fr}
\emailAdd{faisal.karimi@kcl.ac.uk}
\emailAdd{tanmoys@cmi.ac.in}
\emailAdd{adarshsu@imsc.res.in}
\emailAdd{gerard.watts@kcl.ac.uk}

\abstract{
In this paper we make a proposal for the solution to a long-standing problem - the asymptotic expansions of the modular $S$-transform of a generalised Gibbs ensemble (GGE) in a theory with $\CW_3$ symmetry where the GGE includes the first non-trivial charge. Equivalently, we give a proposal for the modular $S$-transform of traces of arbitrary powers of the zero mode $W_0$. We provide evidence in the form of exact results using Zhu's recursion, results obtained using conjectured results for Verma modules, and exact results for the particular value $c=-2$. We expect these have generalisations to other symmetry algebras/hierarchies such as the Virasoro algebra/KdV charges, and to GGEs with arbitrary finite sets of charges.}

\newcommand{\lb}{)^{\vphantom{y}}_2}

\allowdisplaybreaks
\begin{document}

\maketitle

\section{Introduction}

Central to two-dimensional conformal field theory are torus correlation functions,
\begin{align}
\vev{{\,\CO\,}}_{\tau}   
\;,
\label{eq:torus}
\end{align}
where $\tau$ is the modular parameter and $\CO$ is some function of the fields in the conformal field theory (CFT) \cite{DiFrancesco:1997nk}.
In the case $\CO=1$ this is simply the torus partition function which is invariant under the group of modular transformations.

Torus correlation functions \eqref{eq:torus} can be constructed from 
traces over spaces $\CH_i$ of the form
\begin{align}
    \Tr_{\CH_i}( \,\hat{\CO}\,q^{L_0 - c/24})\;,
\label{eq:Trdef}
\end{align}
where $\hat\CO$ is an operator, $q=\exp(2\pi i \tau)$, $c$ is the central charge of the Virasoro algebra and $L_0$ is the element of the algebra that generates dilations on the plane. 
It is well known that in the case 
$\CO=1$, ie $\hat\CO=\identity$ (the identity operator), these functions have well-defined modular properties - in the case of rational conformal field theories with chiral algebra $\CA$ with $\CH_i$ the finite set of representations of $\CA$, these functions form a vector-valued modular form, and this has been a principal tool in the study of conformal field theories since it was first proposed by Cardy \cite{Cardy:1986ie}.
We will often use the term ``thermal correlators" for the traces \eqref{eq:Trdef}.

We are interested in generalisations of the torus correlation functions \eqref{eq:torus} and traces \eqref{eq:Trdef} where the insertion $\CO$ is a power, or an exponential, of a conserved charge.
All conformal field theories contain  infinite sets of commuting conserved charges $\{\hat I_i\}$ \cite{Sasaki:1987mm,Zamolodchikov:1989hfa}, and in such cases it is natural to extend the usual idea of a Gibbs ensemble (corresponding to a statistical ensemble weighted by the energy of a configuration) to a Generalised Gibbs Ensemble (GGE) \cite{Essler:2014qza} with a separate chemical potential for each conserved charge. For GGEs the traces \eqref{eq:Trdef} become
\be
    \Tr_{\CH_i}\left( e^{\sum_i \a_i \hat I_i} q^{L_0 - c/24}\right) \;.
\ee
The properties of these and similar traces are  relevant in many physical situations, such as Black holes \cite{Kraus:2011ds,Gaberdiel:2012yb,Dymarsky:2020tjh}, CFTs in the limit of large central charge \cite{Dymarsky:2018lhf,Dymarsky:2018iwx,Dymarsky:2022dhi}, as well as being of intrinsic mathematical interest \cite{hrj:8932}. 
Recent studies of the modular properties of such traces include \cite{Downing:2023lnp,Downing:2023lop,Downing:2024nfb,Downing:2021mfw,Maloney:2018hdg,hrj:8932}.

In this paper we are interested in the particular case of conformal field theories with the Zamolodchikov $\CW_3$ W-algebra symmetry \cite{Zamolodchikov:1985wn}. Such theories have a particular set of commuting conserved charges known as the quantum Boussinesq charges \cite{Kupershmidt:1989bf,Bazhanov:2001xm}. 
We present here a proposal for the modular properties of the GGE in the presence of the simplest non-trivial charge, 
\begin{align}
    \Tr_{\CH_i}(\, e^{\a W_0}\,q^{L_0 - c/24})
\end{align}
and correspondingly of the traces
\begin{align}
    \Tr_i\left(\, (W_0)^n \,q^{L_0 - c/24}\,\right)
\equiv    \Tr_{\CH_i}\left(\, (W_0)^n \,q^{L_0 - c/24}\,\right)
    \;.
    \label{eq:TrW0n}
\end{align}
where $W_0$ is the zero-mode of the spin-3 generator of the $\CW_3$ algebra and $\CH_i$ is a representation of the $\CW_3$ algebra with known modular properties.
This is a long-standing problem, first posed in 
\cite{Gutperle:2011kf}.
To be precise, we give a solution for the modular properties of the traces 
\eqref{eq:TrW0n}
under the particular modular $S$ transformation
\begin{align}
    \tau \mapsto -1/\tau
    \;,\;\;\;\;
    \hat q = e^{-2\pi i/\tau}
    \mapsto
    q = e^{2 \pi i \tau}
    \;,\;\;\;\;
        \Tr_{i}({\hat q}^{L_0-c/24})
    =
    \sum_j S_{ij} 
    \Tr_{j}({q}^{L_0-c/24})
\;,
\end{align}
in the form of a solution for the operator $\hat Q$ in the generating function,
\begin{align}
    \Tr_{i}(\,e^{\alpha W_0}\,{\hat q}^{L_0-c/24})
    \sim 
    \sum_j S_{ij} 
    \Tr_{j}(\,e^{\hat Q}\,{q}^{L_0-c/24})\;.
    \label{eq:proposal}
\end{align}
Note that \eqref{eq:proposal} is only an equality of asymptotic expansions in $\a$ - in general the exact relation for GGEs involves non-perturbative terms, as shown in \cite{Downing:2023lnp} for KdV charges - but this is the simplest way to express our result for the individual terms \eqref{eq:TrW0n}. One notable result is that there is no special role at all played by the charges of the Boussinesq hierarchy. In the first results for the KdV charges in the free fermion model, the modular transform of the GGE could be expressed in terms of the KdV charges, but this was just an``accident"; it was already shown in \cite{Downing:2024nfb} that this only happens at special values of the central charge; it appears that for the $\CW_3$ GGE, there is no central charge at which the modular transform of the GGE can be expressed in terms of the Boussinesq charges.

The traces \eqref{eq:TrW0n} have already been studied in
\cite{Iles:2013jha,Iles:2014gra,Gutperle:2011kf,Gaberdiel:2012yb}, and partial results for their modular transforms found for $n\leq 3$; in all cases these agree with our proposal. We have further checked our proposal for $n\leq 4$ using the methods of \cite{Gaberdiel:2012yb} and for $n\leq 7$ using the Verma module methods of \cite{Iles:2014gra} and for all $n$ in the particular case of $c=-2$ using a generalisation of the methods of \cite{Downing:2023lop}.

The solution we propose in equation \eqref{eq:conj1b} is clearly related to the work of Dijkgraaf \cite{Dijkgraaf:1996iy} (see \cite{Gui:2025kbd} for a recent treatment), but the particular currents that appear in our expressions do not satisfy the conditions of a pre-algebra that is required for his construction to work, as we show in appendix \ref{app:Dijkgraaf}.

We have singled out the particular traces 
\eqref{eq:TrW0n} for study because of the extensive previous works. We have checked a corresponding proposal for similar traces of powers of the first non-trivial KdV charge, and see no reason why it should not generalised to all combinations of charges, but as we have not yet checked these, we will only mention these again in the conclusions.

The structure of the paper is as follows. 

In section \ref{sec:conjecture}, we discuss the relation between expressions on the torus and the plane and 
make the proposal \eqref{eq:proposal} explicit and in section \ref{sec:check}, we give the perturbative expansion of the proposal and explain the steps needed to check the it perturbatively.
The next three sections are concerned with two ways to calculate the terms in the perturbative expansion.

In section \ref{sec:4}, we use the method of \cite{Gaberdiel:2012yb} to compute the modular transformation of the traces of $n$ zero modes as an integral of an $n$-point correlator. In order to compute the $n$-point correlator, we employ Zhu recursion relation \cite{zhu1990vertex}. These are rigorous calculations but hard to carry out to high order. 

In section \ref{sec:mldo}, 
we explain how the trace over any particular combination of modes of fields can be expressed as a modular linear differential operator (MLDO) acting on the particular traces \eqref{eq:TrW0n}. In section \ref{sec:Verma} we show how these traces take a particularly simple form in the case of Verma module representations and use the methods of \cite{Iles:2014gra} to find the modular properties of the traces \eqref{eq:TrW0n} for $n \leq 7$ and find agreement with our proposal. 
As with the Zhu recursion method, these give results valid for any modular invariant set of representations but rely on various assumptions on the properties of traces and Verma modules. The advantage is that they can be carried out to rather higher order in $\a$.

 In section \ref{sec:c=-2}, we show that the proposal is actually true in the particular case of the $\CW_3$ algebra as it arises in the theory of symplectic fermions at $c=-2$. This also goes beyond a perturbative or asymptotic treatment to give an exact result in $\a$.
 
 In the conclusion, we discuss possible extensions of our proposal. Finally, we include several appendices with technical details that would otherwise obscure the text.

\section{Conjecture}
\label{sec:conjecture}

The main result of this paper is a conjecture for the form of the charge $\hat Q$ appearing in the asymptotic expansion of the modular transform of the GGE with a single $W_0$ mode inserted,
\begin{equation}
    \Tr_i(
    \,e^{\a W_0}\,\hat{q}^{L_0-c/24}) \sim 
    \sum_j \,S_{ij}\,\Tr_{j}(\,e^{\hat Q}\,q^{L_0-c/24}).
      \label{eq:conj1-0}
\end{equation}
We can think of this asymptotic equality as following from a relation between torus correlation functions,
\begin{align}
    \left\langle
    \, \exp\left({\alpha\textstyle\int} W\, \frac{\rd x}{2\pi}\right)\,\right\rangle_{-1/\tau}
    \sim \left\langle
    \, \exp\left(\alpha\tau^3{\textstyle\int} \CW \, \frac{\rd x}{2\pi}\right)\,
    \right\rangle_\tau
    \;.
    \label{eq:torusconjecture}
\end{align}
Our proposal is that the field $\CW$ is defined (up to total derivatives) by 
\begin{align}
    \CW = 
    \sum_{n=0}^\infty
    \frac{1}{n!}
    \left( \frac{\alpha\tau^2}{4\pi i}\right)^n
    [W^{n+1}]
    \;.
    \label{eq:conj1a}
\end{align}
The fields $[W^n]$ have conformal weight $n+2$ and can be defined recursively as follows,
\begin{align}
    [W^n] = \sum_{m=1}^{n-1}
    \,\binom{{n-2}}{{m-1}}\,( \,[W^m]\, [W^{n-m}] \,\lb
    \;,\;\;\;\;
    [W] = W
    \;,\label{eq:conj1b}
\end{align}
where $(AB\lb$ is the second product structure defined in \cite{Dijkgraaf:1996iy}, 
and is the coefficient of the second order pole in the OPE $A(z)B(w)$ expanded about $w$. The product $(AB\lb$ is commutative (up to total derivatives) but not associative. We give some more details of the construction of $[W^m]$ and explicit examples of these fields in appendix \ref{app:otter}.

The result of \eqref{eq:torusconjecture}
is that the $\hat Q =  \a\t^3\hat\CW_0$ and so traces on the plane satisfy
\begin{equation}
    \Tr_i(
    \,e^{\a W_0}\,\hat{q}^{L_0-c/24}) \sim 
    \sum_j \,S_{ij}\,\Tr_{j}(\,e^{\alpha\tau^3\hat{\CW}_0}\,q^{L_0-c/24}).
      \label{eq:conj1}
\end{equation}
where 
$\hat\CW_0$ is the zero-mode of 
the field $\hat\CW$, defined in terms of the map of $\CW$ to the plane in appendix \ref{app:Gaberdiel}.

We do not (as yet) have a general proof of the conjecture \eqref{eq:conj1a}, \eqref{eq:conj1b} and \eqref{eq:conj1}, but have instead checked it extensively in different regimes with various different methods, and can prove that it is correct in the case of the symplectic fermion construction at $c=-2$.

It is important to note that the conjecture for the forms of the asymptotic expansion 
\eqref{eq:conj1}
is simplest when expressed in terms of correlation functions on the torus \eqref{eq:torusconjecture}, since it is the complex structure on the torus which is invariant under modular transformations. As a simple example, consider the first non-trivial term in the expansion \eqref{eq:conj1a} of $\CW$, namely $[W^2]$. From the operator product expansion \eqref{eq:WWope}, we find that the field $[W^2]$ is
\begin{align}
    [W^2](z) = \frac23 \Lambda(z) + 
    \text{(total derivatives)}
\end{align}
The contribution to \eqref{eq:torusconjecture} is (up to an overall factor)
\begin{align}
    \left\langle\,
    {\textstyle\int_0^{2\pi i}}[W^2](u)\,\frac{\rd u}{2\pi i}\,\right\rangle_\tau
    = 
    \frac 23\left\langle\,
    {\textstyle\int_0^{2\pi i}}\Lambda(u)\,
    \frac{\rd u}{2\pi i}\,\right\rangle_\tau
    =\frac 23 \Tr\left( \,
    \hat\Lambda_0\,\,q^{\Ione}\,\right)
    \;.
\end{align}
Under the
map from the cylinder to the plane 
$u\mapsto z = e^{u}$, $\Lambda(u)$ transforms as  \cite{Gaberdiel:1994fs},
\begin{align}\label{eq:Lambda cyl to plane}
    \hat\Lambda(u) \mapsto 
    z^4\Lambda(z) - \frac{5c+22}{60}z^2T(z) + \frac{c(5c+22)}{2880}\;,
\end{align}
and its integral maps to 
the first non-trivial KdV charge, which we denote $\hat\Lambda_0$ (see appendix \ref{app:Gaberdiel} for our conventions),
\begin{align}
    \hat I_3^{\text{KdV}} = \hat\Lambda_0 = \Lambda_0 - \frac{5c+22}{60}L_0 + \frac{c(5c+22)}{2880}
    \;.
\end{align}
The corresponding traces are (with $\SD n$ being the Serre derivatives -- see appendix \ref{app:MF}),
\begin{align}
    \Tr( \hat\Lambda_0\,q^{\Ione})
    &= \left(\,
    \SD{2}\SD{0} 
    + \frac{c E_4}{1440}
    \,\right)\Tr(q^{\Ione})\;,
    \\
    \Tr( \Lambda_0\,q^{\Ione})
    &=
    \left(\,
    \SD{2}\SD{0} 
    + \frac{c E_4}{1440} 
    + \frac{5c+22}{60}\SD{0} 
    + \frac{c(5c+22)}{2880}
    \,\right)
    \Tr(q^{\Ione})\;.
\end{align}
The trace over $\hat\Lambda_0$, coming from the torus correlation function of the integral of a field of definite conformal weight, is a modular form (in this case of weight 4); the trace over $\Lambda_0$ has no particular modular properties. This is an example of why modular properties are best expressed in terms of fields on the torus, and the corresponding results for thermal correlation functions require the map from the cylinder to the plane to have simple expressions.

The remainder of the paper is devoted to explaining the checks of our conjecture as outlined in the introduction.

\section{Perturbative test of the conjecture}
\label{sec:check}

In order to test the conjecture \eqref{eq:conj1} perturbatively,  we need to expand both sides as power series in $\a$ and check to as high an order as is feasible that they agree.

The expansion of the left hand side of \eqref{eq:conj1} is easy,
\begin{align}
    \Tr_i(e^{\a W_0} \hat q^{\Ione})
    = \sum_n \frac{\a^n}{n!} \Tr_i(W_0^n\,\qhat^{\Ione})
    \;.
\end{align}
We calculate the modular transforms of these traces exactly for $n\leq 4$ in section \ref{sec:4}, and for $n\leq 8$ assuming various conjectures in section \ref{sec:vermamodular2}. The results are sums of modular linear differential operators (MLDOs) acting on the functions $\Tr_i(W_0^m\,q^{\Ione})$; we give the results for $n=2$ in \eqref{eq:WW}, $n=3$ in \eqref{eq:W0cube} and $n=4$ in \eqref{eq:W0fourth}. We do not give the expressions for $n>4$ as they are rather lengthy.

The right hand side of \eqref{eq:conj1} is
\begin{align}
    \Tr_j(e^{\a\tau^3\hat \CW_0}\,q^{\Ione})
    \;,
    \label{eq:ExpaWhat}
\end{align}
where $\hat\CW$ is itself a series in $\a$, \eqref{eq:conj1a},
\begin{align}
  \a\t^3\hat\CW &= 
    \a\t^3\sum_{n=0}^\infty
    \frac{1}{n!}
    \left( \frac{\alpha\tau^2}{4\pi i}\right)^n
    [\hat W^{n+1}]
    \nonumber\\
    &=
    \a\t^3W
+ \frac{\a^2\t^5}{4\pi i}
         [\hat W^2]
+ \frac 12 \frac{\a^3\t^7}{(4\pi i)^2}
         [\hat W^3]
+ \frac 16 \frac{\a^4\t^9}{(4\pi i)^3}              [\hat W^4] + \ldots
    \;.
    \label{eq:conj1aprimed}
\end{align}
Expanding \eqref{eq:ExpaWhat} in $\a$ we then find
\begin{align}
    &\Tr_j(e^{\a\tau^3\hat \CW_0}q^{\Ione})
    \nonumber\\
    =&
\Tr_j(q^{\Ione})    
+\a
\Tr_j\left(\left(\t^3 W_0\right)q^{\Ione}\right)
+{\a^2}
\Tr_j\left(\left(
\frac {\t^6}2 W_0^2
+ 
\frac{\t^5}{4\pi i}[\hat W^2]_0
\right)q^{\Ione}\right)\;+
\nonumber\\
&+{\a^3}
\Tr_j\left(\left(
\frac {\t^9}6   W_0^3
+
\frac{\t^8}{4\pi i} W_0\,[\hat W^2]_0
+
\frac 12\frac{\t^7}{(4\pi i)^2} [\hat W^3]_0
\right)q^{\Ione}\right)
\nonumber\\
&+{\a^4}
\Tr_j\left(\left(
\frac{\t^{12}}{24} W_0^4
+
\frac{\t^{11}}{8\pi i}W_0^2\,[\hat W^2]_0
-
\frac{\t^{10}}{32 \pi^2}\,([\hat W^2]_0^2 {+} [\hat W^3]_0)
+
\frac 16 \frac{\t^9}{(4\pi i)^3}[\hat W^4]_0
\right)q^{\Ione}\right)
\nonumber\\
&+\ldots
\nonumber\\
&+\a^6\Tr_j\left(
\left(
\ldots
- 
\frac{\t^{16}}{384 \pi^2}
\left(
4 (W_0)^2\,([\hat W^2]_0)^2
+ 
2 W_0\,[\hat W^2]_0\,W_0\,[\hat W^2]_0
\right) 
+ \ldots \right)q^{\Ione} 
\right)
\nonumber\\
&+\ldots
\label{eq:expand}
\end{align}
These include traces of products of the zero modes $[\hat W^m]_0$ -- see appendix \ref{app:otter} for details of the fields $[\hat W^m]$. As explained in section \ref{sec:mldo}, the terms in \eqref{eq:expand} can also be expressed as sums of modular linear differential operators acting on the traces $\Tr_i(W_0^m q^{\Ione})$. 
Note the appearance at order $\a^6$ of the first term which includes the symmetrisation of the four-fold product of two non-commuting zero modes; the two separate terms are not quasi-modular forms of definite weight, but the symmetrised term is, as shown in section \ref{sec:4terms}.

To calculate the modular linear differential operators for the various traces in \eqref{eq:expand}, we can use either the exact method of \cite{Iles:2013jha} or the  method of section \ref{sec:mldo} which relies on assuming various properties of the functions in question.

Having expressions for the both sides of \eqref{eq:conj1} allows
allows us to check the correctness of our conjecture at any order in $\a$. 
As an example, we can check the term at order $\a^3$ using the results 
\eqref{eq:W0cube} for the left-hand-side and  
\eqref{eq:TrW3} and
\eqref{eq:TrW0W20} for the right-hand-side.
We find complete agreement between the expansions of the two sides of \eqref{eq:conj1} as far as we have been able to check (which is currently order $\a^7$). 

In the next few sections we explain how to calculate the terms in this perturbative expansions.

\section{Modular transformation of the zero modes}
\label{sec:4}

In this section, we make use of a first-principles method from \cite{Gaberdiel:2012yb} that allows us to arrive at the S-transforms of traces involving products of zero modes. 
Unlike the subsequent sections \ref{sec:mldo} and \ref{sec:Verma}, the results of this section are rigorously proven, assuming the standard properties of conformal field theory.

Our goal is to compute the modular transform of the GGE in the presence of a single conserved charge $W_0$ of the form 
\begin{equation}
    \Tr_i \left( q^{L_0-\frac{c}{24}} e^{\alpha W_0} \right) \;. 
\end{equation}
In order to do so, we first take the asymptotic expansion of the above in the chemical potential $\alpha$ and get
\begin{equation}
 \sum_{n=0}^{\infty}\Tr_i \left( q^{L_0-\frac{c}{24}}\frac{\alpha^n(W_0)^n}{n!} \right).    
\end{equation}
Next, we perform the modular transformation of each term in the asymptotic expansion.\\ This section is devoted to the computation of modular transformation of the individual terms in the above asymptotic expansion. The discussion here follows closely that of Section 2.3 from \cite{Gaberdiel:2012yb}.

Using the convention of \cite{Gaberdiel:2012yb}, we define 
the contribution from the representation $r$ to 
the  n-point correlator on the torus as 
\begin{equation}
        F_r((a^1,z_1) \cdots (a^n,z_n) ; \tau) = \text{Tr}_r\,[ V(a^1,z_1) \cdots V(a^n,z_n) q^{L_0-\frac{c}{24}}] \prod_{i=1}^n z^{\wgt(a^i)}~,
\end{equation}
where the trace is over any representation $r$ of the vertex algebra, $z_i=e^{2\pi i u_i}$ where $u_i$ are the coordinates on the torus\footnote{In this section we take the torus to the complex plane modulo translations by 1 and $\tau$}, 
$V(a,z)$ is the vertex operator corresponding to the state $a$ on the plane,
and $\wgt(a^i)$ correspondence to the conformal dimension of the state $a^i$ whereas $(a^i,z_i)$ is a field on the cylinder corresponding to the state $a^i$.

For Virasoro primaries $a^i$, the modular transformation property of the n-point correlator is given by:
\begin{equation}  
       F_r\left(\!(a^1,z_1) \cdots (a^n,z_n) \, ;\frac{a\tau{+}b}{c\tau{+}d} \right) = (c\tau{+}d)^{\sum_j \wgt(a^j)} \sum_s M_{rs} F_s\left( (a^1,z_1^{    c\tau+d}) \cdots (a^n,z_n^{c\tau+d}) ; \tau \right)\label{Frmodtransform}
\end{equation}
where $M_{rs}$ is a representation of the modular group. Recall that our interest is in the modular transformation of zero-mode insertions of the form: 
\begin{equation}
    \Tr_r \left( W_0^n q^{L_0-\frac{c}{24}} \right)~.
\end{equation}
These can be obtained from \eqref{Frmodtransform} by performing the integrals  $ \frac{1}{2\pi i}\oint \frac{dz_i}{z_i}$, which turn the LHS  of \eqref{Frmodtransform} into the S-transformed trace of the zero modes of the respective local fields. Changing variables to go to $z'_i = z_i^\tau$, we find\footnote{Strictly speaking, the integration contours are all displaced slightly by phases to avoid the divergences from overlapping contours. However, the final answers to the integrals is independent of these phases \cite{Gaberdiel:2012yb}. }:
\begin{equation}
\label{eq:GabStrans}
\begin{split}
        &F_r\left( a^1_0 \cdots a^n_0 \, ; -\frac{1}{\tau} \right) \\
        =&{}\;\frac{1}{(2\pi i)^n} \int_1^q \frac{dz'_1}{z'_1} \, \cdots \int_1^q \frac{dz'_n}{z'_n} \, \tau^{-n+\sum_j \wgt(a^j)} \, \sum_s S_{rs} \,  F_s\left( (a^1,z_1) \cdots (a^n,z_n) \, ; \tau \right)
        \end{split}
\end{equation}
The n-point function on the right hand side of \eqref{eq:GabStrans} can be calculated using Zhu recursion relation \cite{zhu1990vertex}, which we will discuss later in the section.

In the remainder of this section, we will explicitly compute the $S$ transformation of the products of zero mode of the spin three field $W$, which is relevant to compute the $S$ transformed Generalized Gibbs ensemble.

\subsection{Correlators from Zhu Recursion}
To compute the $n$-point correlator on the right-hand side of equation \eqref{eq:GabStrans}, we employ Zhu’s recursion relation\footnote{A concise overview of Zhu’s recursion relation is provided in Appendix \ref{app:Zhu}. For a comprehensive treatment, we refer the reader to \cite{Ashok:2024zmw,Gaberdiel:2012yb}}. The key idea of this recursion is as follows: given an $n$-point function, we first expand the first field in terms of its modes. Retaining only the zero-mode contribution of this field, we commute the remaining non-zero modes through the other fields and use the cyclic property of the trace to return them to their original position. This procedure allows us to express the $n$-point correlator as a sum of two contributions: an $(n-1)$-point correlator, where the first field is replaced by its zero-mode insertion, and a set of $(n-1)$-point correlators in which one of the remaining fields is replaced by the result of the action of the non-zero modes of the first field on that field. By iterating this process, we recursively reduce the $n$-point function to a sum of traces involving only zero-mode insertions. Zhu’s theorem guarantees that this recursion terminates, ultimately expressing the original correlator purely in terms of zero-mode expectation values. Since the Zhu formula can take us from the correlator on the RHS to traces involving only zero modes (with coefficients being quasi-elliptic), performing these integrals reduces the S-transformed trace to an expression involving the non S-transformed traces of products of zero modes. 
\subsubsection{Single zero mode insertion}
Let us first work out the simplest possible case where we will compute the $S$-transformation of a single zero mode insertion. Using \eqref{eq:GabStrans}, we can write
\begin{equation}
\label{eq:oneW1}
      F_r\left( W_0  ; -\frac{1}{\tau} \right) = \frac{1}{(2\pi i)}\int_1^q \frac{dz'_1}{z'_1}  \tau^{2} \, \sum_s S_{rs} \,  F_s\left( (W,z_1) ; \tau \right).
\end{equation}
We can compute the thermal one point function on the right hand side using the Zhu recursion relation \eqref{eq:zhu} and get 
\begin{equation}
  F_s\left( (W,z_1), ; \tau \right)=\Tr_s\left( W_0  q^{L_{0}-\frac{c}{24}} \right).
\end{equation}
We can write \eqref{eq:oneW1} as 
\begin{equation}
  F_r\left( W_0  ; -\frac{1}{\tau} \right) =  \tau^{3} \, \sum_s S_{rs} \,  \Tr_s\left( W_0  q^{L_{0}-\frac{c}{24}} \right) \;,
\end{equation}
Hence the thermal correlator transforms as a weight 3 modular form, as expected.

\subsubsection{Two zero mode insertion}
Let us now compute the $S$ transform of the two zero mode insertion. From \eqref{eq:GabStrans}, we can write
\begin{equation}
\label{eq:oneWW1}
      F_r\left( W^2_0  ; -\frac{1}{\tau} \right) = \frac{1}{(2\pi i)^2}\int_1^q \frac{dz'_1}{z'_1}\int_1^q \frac{dz'_2}{z'_2}  \tau^{4} \, \sum_s S_{rs} \,  F_s\left( (W,z_1), (W,z_2); \tau \right).
\end{equation}
We compute the two-point function\footnote{Details of the computation can be found in the appendix B of \cite{Ashok:2024zmw}.} $F_s\left( (W,z_1), (W,z_2); \tau \right)$ using the Zhu recursion and get
\vspace{0.2 cm}
\begin{equation}
\begin{aligned}
\label{eq:WW}
&F_s((W,z_1),(W,z_2);\tau)
\\=&\Tr_s\left( (W_0)^2  q^{L_{0}-\frac{c}{24}} \right)+\frac{1}{(2\pi i)^{2}}\mathcal{P}_{2}(z_{21})\Big(\frac{2}{3}\partial^{2}\chi_s(\tau)-\frac{1}{9}\partial \chi_s(\tau)+\frac{c}{2160}\chi_s(\tau)\Big) \\
&+\frac{1}{(2\pi i)^{2}}\mathcal{P}_{2}(z_{21})\frac{4}{3}\Big(\frac{1-E_{2}(\tau)}{12}\partial \chi_s(\tau)+c\frac{E_{4}(\tau)-1}{2880}\chi_s(\tau)\Big) \\
&+\frac{2}{3b^2(2\pi i)^{4}}\mathcal{P}_{4}(z_{21})\partial \chi_s(\tau)+\frac{c}{9b^2(2\pi i)^{6}}\mathcal{P}_{6}(z_{21})\chi_s(\tau)~,
\end{aligned}
\end{equation}
where $z_{ij}=\frac{z_{i}}{z_j}$, $\partial=q\frac{\partial}{\partial q}$ and $\mathcal{P}_{k}$ is the  Weierstrass function. Next we perform the integrals using the list of integrals given in appendix \ref{app:c} and get\footnote{Note that our answer doesn't exactly match with \cite{Iles:2014gra} as the $\CW_3$ algebra is different in convention.}
\vspace{0.2 cm}
\begin{equation}
\label{eq:W0sq}
 F_r\left( W^2_0  ; -\frac{1}{\tau} \right)=\sum_s S_{rs}\Big\{\tau^6  \Tr_s\left( W^2_0  ; \tau \right)+\frac{\tau^{5}}{3\pi i}\Big[D^{(2)}D^{(0)}+\frac{cE_4}{1440}\Big]\chi_s(q)\Big\},
\end{equation}
where $\SD k$ are the Serre derivatives. Here $\chi_s$ is the reduced character of the $\CW_3$ algebra, defined as 
\begin{equation}
 \chi_{s} = \text{Tr}_s\left( q^{L_0-\frac{c}{24}}\right)~,
\end{equation}
where the trace is taken over some representation of the Hilbert space of the $\CW_3$ algebra and $\Tr_s\left( W^n_0  ; \tau \right)$ is defined as 
\begin{equation}
    \Tr_s\left( W^n_0  ; \tau \right)=\Tr_s\left( W^2_0\, q^{L_0-\frac{c}{24}}\right)
\end{equation}

\subsubsection{Higher zero mode insertion}
The same computational strategy can be extended to derive the $S$-transformation for three or more zero modes. However, due to the highly cumbersome nature of the resulting expressions, we restrict our presentation here to reporting only the final outcome.
\begin{equation}
\begin{aligned}
\label{eq:W0cube}
 F_r\left( W^3_0  ; -\frac{1}{\tau} \right)&=
 \sum_s S_{rs}\Big\{\tau^9  \Tr_s(W^3_0;\tau)+\frac{\tau^{7}}{(2\pi i)^2}6D^{(3)}\Tr_s(W_0;\tau)\\&~~~~~~~~~~+\frac{\tau^{8}}{(2\pi i)} \Big[2D^{(5)}D^{(3)}+E_2(\tau)D^{(3)}+\frac{(c+30)E_4(\tau)}{720}\Big]\Tr_s(W_0;\tau)\Big\}
\end{aligned}
\end{equation}
\small{
\begin{equation}
\begin{aligned}
\label{eq:W0fourth}
 F_r\left( W^4_0  ; -\frac{1}{\tau} \right)&=
 \sum_s S_{rs}\Big\{\tau^{12}  \Tr_s(W^4_0;\tau)+\\&\frac{\tau^{11}}{(2\pi i)}\Big[\Big(\frac{(c+60) E_4(\tau)+180E_2^2(\tau)}{360} + 4D^{(8)}D^{(6)}+4E_{2}(\tau)D^{(6)}\Big)\Tr_s(W^2_0;\tau)+\\&
 \Bigg(\frac{\left(-E_2^3(\tau)+3 E_{2}(\tau) E_{4}(\tau)-2 E_{6}(\tau)\right)}{81}D^{(2)}D^{(0)}+\\
 &\frac{(5 c+22) \big(E_2^2(\tau) E_4(\tau)-2 E_2(\tau) E_{6}(\tau)+E_{4}^2(\tau)\big)}{7776} D^{(0)}-\frac{c \bigg((5 c+22) E_2^2(\tau) E_{6}(\tau)}{933120}-
 \\&\frac{2 (5 c+34) E_2(\tau) E_4^2(\tau)+(5 c{+}38) E_{4}(\tau) E_{6}(\tau)+8 E_2^3(\tau) E_4(\tau)\bigg)}{933120}\Bigg)\chi_s(\tau)\Big]+\frac{\tau^{10}}{(2\pi i)^2}\\&\Big[\Big(24D^{(6)}+6E_{2}(\tau)\Big)\Tr_s(W^2_0;\tau)+\Bigg(\frac{4}{3}D^{(6)}D^{(4)}D^{(2)}D^{(0)}+\frac{8}{9}E_2(\tau)D^{(4)}D^{(2)}D^{(0)}+\\&\Big(\frac{1}{540} (c+160) E_4(\tau)-\frac{2 E_2^2(\tau)}{9}\Big)D^{(2)}D^{(0)}+\frac{(27 c+110) (E_2(\tau)E_4(\tau)-E_6(\tau))}{3240}D^{(0)}\\+& \Big(\frac{c \left(-20 (5 c+38) E_{2}(\tau) E_{6}(\tau)+(101 c+1000)E_4^2(\tau)-240 E_2^2(\tau) E_4(\tau)\right)}{1555200}\Big)\Bigg)\chi_s(\tau)\Big]\\+&\frac{\tau^{9}}{(2\pi i)^3}\Big[24\Tr_s(W^2_0;\tau)+\Bigg(\frac{80}{9}D^{(4)}D^{(2)}D^{(0)}-\frac{4}{3}E_{2}(\tau)D^{(2)}D^{(0)}\\&+\frac{1}{81} (3 c+11) E_4(\tau)D^{(0)}-\frac{c  ((5 c+62) E_6(\tau)+18 E_{2}(\tau)E_4(\tau))}{19440}\Bigg)\chi_s(\tau)\Big]\Big\}
\end{aligned}
\end{equation}}
\normalsize
The introduction of derivatives of Eisenstein series allows for a partial reduction of this expression, as demonstrated below. However, it remains unclear whether a ``simplest" form exists or if this reduction is merely a contingent phenomenon.
\begin{align}
\label{eq:W0fourth2}
 &F_r\left( W^4_0  ; -\frac{1}{\tau} \right)
 \nonumber\\
 &=
 \sum_s
 S_{rs}\Big\{\tau^{12}  \Tr_s(W^4_0;\tau)\nonumber
 \\&
 +\frac{\tau^{11}}{(2\pi i)}\Big[\Big(
 \frac{c}{360}E_4 
 - 4 E_2'
 -\frac 83 E_2 D^{(0)}
 + 4 D^{(0)}D^{(0)}
 \Big)\Tr_s(W^2_0;\tau)\nonumber
 \\&\qquad\qquad+\Bigg( - \frac 89 E_2''D^{(2)}D^{(0)}
 +\frac{(5 c+22)}{1080} E_4'' D^{(0)}
 - \frac{c(5c+46)}{272160} E_6''
 - \frac{c}{8100} E_4'''
 \Bigg)\chi_s(\tau)\Big]
\nonumber \\&+\frac{\tau^{10}}{(2\pi i)^2}\Big[\Big(24D^{(6)}+6E_{2}\Big)\Tr_s(W^2_0;\tau)+\Bigg(\frac{4}{3}D^{(6)}D^{(4)}D^{(2)}D^{(0)}+\frac{8}{9}E_2D^{(4)}D^{(2)}D^{(0)}
\nonumber \\&\qquad\qquad
 +\Big(\frac{(c+160)}{540} E_4-\frac{2}9 E_2^2\Big)D^{(2)}D^{(0)}
 +\frac{(27 c+110)E_4'}{1080}D^{(0)}
 \\&\qquad\qquad
 + \Big(\frac{c \left(-20 (5 c+38) E_{2} E_{6}
 +(101 c+1000)E_4^2
 -240 E_2^2E_4\right)}{1555200}\Big)\Bigg)\chi_s(\tau)\Big]\nonumber
 \\&+\frac{\tau^{9}}{(2\pi i)^3}\Big[24\Tr_s(W^2_0;\tau)
 +\Bigg(\frac{80}{9}D^{(4)}D^{(2)}D^{(0)}-\frac{4}{3}E_{2}D^{(2)}D^{(0)}+\frac{(3c+11)}{81} E_4 D^{(0)}
\nonumber \\&\qquad\qquad-\frac{c  ((5 c+62) E_6+18 E_{2}E_4)}{19440}\Bigg)\chi_s(\tau)\Big]\Big\}\nonumber
\end{align}

Note that here the primes are shorthand for  $E_n'=\SD{0}E_n$, $E_n''=(\SD{0})^2 E_n$. etc.

\section{Modular Linear Differential Operators}
\label{sec:mldo}

In this section we show how the thermal correlators of the charges that arise in the $\CW_3$ algebra can be written as modular linear differential operators (MLDOs) acting on the thermal correlators $\Tr(W_0^n q^{\Ione})$. These expressions will only depend on the central charge $c$ and are valid for any representation that we trace over since they can be derived by using the cyclicity of the trace and commuting modes within the trace only. These results do not give expressions for $\Tr(W_0^n q^{\Ione})$, only the other thermal correlators in terms of $\Tr(W_0^n q^{\Ione})$. 
This is a necessary step ahead of checking the conjecture in section \ref{sec:check} using the results from section \ref{sec:Verma}.

In order to compute the modular transform of $\Tr(W_0^n q^{\Ione})$ we will choose to take the trace over the Verma module representations defined in appendix \ref{sec:Wreps}, which we discuss in the next section.

\subsection{MLDOs for generic representations}\label{sec:MLDO generic reps}

One way to compute the modular transformation of a given thermal correlator is to assume that it can be written as a modular linear differential operator acting on the thermal correlators $\Tr(W_0^n q^{\Ione})$. In the case where the charges are constructed out of just the Virasoro modes $L_n$ this method was used in \cite{Maloney:2018hdg} to construct the thermal correlators of KdV charges as modular differential equations acting on just the character $\chi = \Tr(q^{\Ione})$. We will extend the argument in \cite{Maloney:2018hdg} to the charges found in $\CW_3$ models.

We are interested in charges that are the zero modes of fields. These charges are the zero mode of fields $\hat J(u)$ which are the image of the field $J(u)$ when it is mapped from the cylinder to the plane under the map $u \to z = e^{iu}$, as in \eqref{eq:Lambda cyl to plane}. 

We note that at each level we have multiple fields and therefore multiple charges. Let $\hat I_{n-1;a}$ be a charge that has come from a weight $n$ field. The index $a$ labels the charges that come from different fields with the same weight. Then in \cite{Dijkgraaf:1996iy} it was argued that the thermal one-point function
\be
    \Tr\left(\hat I_{n-1;a} q^{L_0 - \frac{c}{24}}\right) \;,
\ee
is a weight $n$ modular form (see \eqref{eq:MF def} for the definition of modular forms). Furthermore in \cite{Dijkgraaf:1996iy} it was argued if we have $K$ such charges, $\hat I_{n_i-1;a_i}$ for $i=1,\dots,K$ then the thermal $K$-point function
\be\label{eq:I pnt thermal corr}
    \Tr\left(\sum_{\sigma\in S_K} \hat I_{\sigma(n_1-1;a_1)} \dots \hat I_{\sigma(n_K-1;a_I)} q^{L_0 - \frac{c}{24}}\right) \;,
\ee
is a weight $n_1 + \dots + n_K$, depth $K-1$ quasi-modular form (see \eqref{eq:QMF def} for the definition of quasi-modular forms). The sum is over the permutations of the indices of the charges so we take the fully symmetric combination of the charges in the trace. This is required for the correlator to have the correct modular properties.

As concrete examples we will take the correlators with two and three charges. For two charges in the trace we have
\be\begin{split}
    &\Tr\left(\sum_{\sigma\in S_2} \hat I_{\sigma(n_1-1;a_1)} \hat I_{\sigma(n_2-1;a_2)} q^{L_0 - \frac{c}{24}}\right) \\
    &= \Tr\left( \hat I_{n_1-1;a_1} \hat I_{n_2-1;a_2} q^{L_0 - \frac{c}{24}}\right) + \Tr\left( \hat I_{n_2-1;a_2} \hat I_{n_1-1;a_1} q^{L_0 - \frac{c}{24}}\right)\\
    &= 2\Tr\left( \hat I_{n_1-1;a_1} \hat I_{n_2-1;a_2} q^{L_0 - \frac{c}{24}}\right) \;.
\end{split}\ee
The final equality holds using the cyclicity of the trace and the fact that the zero mode of any local field commutes with $L_0$ (and therefore $q^{L_0 - \frac{c}{24}}$). We now move on to traces containing three charges. The thermal correlator that transforms as a quasi-modular form is
\be\begin{split}
    &\Tr\left(\sum_{\sigma\in S_3} \hat I_{\sigma(n_1-1;a_1)} \hat I_{\sigma(n_2-1;a_2)} \hat I_{\sigma(n_3-1;a_3)} q^{L_0 - \frac{c}{24}}\right) \\
    &= \Tr\left( \hat I_{n_1-1;a_1} \hat I_{n_2-1;a_2} \hat I_{n_3-1;a_3} q^{L_0 - \frac{c}{24}}\right) + \Tr\left( \hat I_{n_2-1;a_2} \hat I_{n_3-1;a_3} \hat I_{n_1-1;a_1} q^{L_0 - \frac{c}{24}}\right)\\
    &+ \Tr\left( \hat I_{n_3-1;a_3} \hat I_{n_1-1;a_1} \hat I_{n_2-1;a_2} q^{L_0 - \frac{c}{24}}\right) + \Tr\left( \hat I_{n_1-1;a_1} \hat I_{n_3-1;a_3} \hat I_{n_2-1;a_2} q^{L_0 - \frac{c}{24}}\right)\\
    &+ \Tr\left( \hat I_{n_2-1;a_2} \hat I_{n_1-1;a_1} \hat I_{n_3-1;a_3} q^{L_0 - \frac{c}{24}}\right) + \Tr\left( \hat I_{n_3-1;a_3} \hat I_{n_2-1;a_2} \hat I_{n_1-1;a_1} q^{L_0 - \frac{c}{24}}\right)\\
    &= 3\Tr\left( \hat I_{n_1-1;a_1} \hat I_{n_2-1;a_2} \hat I_{n_3-1;a_3} q^{L_0 - \frac{c}{24}}\right) + 3\Tr\left( \hat I_{n_1-1;a_1} \hat I_{n_3-1;a_3} \hat I_{n_2-1;a_2} q^{L_0 - \frac{c}{24}}\right) \;.
\end{split}\ee
Again the final equality comes from the cyclicity of the trace and the fact the $\hat I_{n_i-1,a_i}$ commute with $L_0$.

In \cite{Dijkgraaf:1996iy} the modular properties of the thermal correlators are derived by swapping the line integrals that define the currents, with surface integrals over the whole torus. Since we are integrating the currents over a surface there is no natural ordering, and therefore we have to take a symmetric combination of the charges in the trace in order to find good modular properties, as we see in the example in section \ref{sec:4terms}.

We can argue similarly to \cite{Maloney:2018hdg} that these thermal correlators can be written as modular linear differential operators (MLDOs) acting this time on $\Tr(W_0^n q^{\Ione})$. The MLDOs are constructed out of Eisenstein series and Serre derivatives (see appendix \ref{app:MF} for definitions) and hence increase the weight of the modular form they act on by an even integer. Therefore if we are constructing an even weighted correlator we must act on $\Tr(W_0^{2n}q^{L_0 - \frac{c}{24}})$ and if we want an odd weighted correlator we must act on $\Tr(W_0^{2n+1}q^{L_0 - \frac{c}{24}})$. 

Considering the odd and the even cases separately we have the following ansatz for the thermal correlators. If $n_1 + \dots + n_K = 2(3N + n)$ with $N=0,1,2,\dots$ and $n = 0,1,2$ then we have
\be\label{eq:even weight MLDO}
    \Tr\!\!\left(\sum_{\sigma\in S_K} \hat I_{\sigma(n_1-1;a_1)} \dots \hat I_{\sigma(n_K-1;a_K)} q^{L_0 - \frac{c}{24}}\!\!\right)\!\! {=} \sum_{l=0}^{N} \sum_{k=0}^{3(N - l) + n} F^{(l)}_{2k,k} D^{3(N - l) + n - k} \Tr \!\!\left(\! W_0^{2l} q^{L_0 - \frac{c}{24}}\!\right) ,
\ee
where $F^{(l)}_{2k,k}$ are quasi-modular forms of weight $2k$ and depth at most $k$ and $D^M = D^{(2M-2+m)} D^{(2n-4+m)} \dots D^{(m)}$ where $m$ is the weight of the modular form being acted on and $\SD k$ are Serre derivatives (see appendix \ref{app:MF} for the definition). 

If instead $n_1 + \dots + n_K = 2(3N + n) - 3 $ with $N=0,1,2,\dots$ and $n = 0,1,2$ then we have
\be\label{eq:odd weight MLDO}
    \Tr\!\!\left(\sum_{\sigma\in S_I} \hat I_{\sigma(n_1-1;a_1)} \dots \hat I_{\sigma(n_I-1;a_I)} q^{L_0 - \frac{c}{24}} \!\!\right)\!\! {=} \sum_{l=1}^{N} \sum_{k=0}^{3(N - l) + n} F^{(l)}_{2k,k} D^{3N + n - 3l - k} \Tr\!\!\left(\! W_0^{2l-1} q^{L_0 - \frac{c}{24}}\!\right) ,
\ee
where again $F^{(l)}_{2k,k}$ are quasi-modular forms of weight $2k$ and depth at most $k$. Note that the $F^{(l)}_{2k,k}$ are different for different correlators. These expressions are valid for any representation we trace over and the quasi-modular forms $F^{(l)}_{2k,k}$ only depend on the central charge $c$.

As explained in appendix \ref{app:MF}, the $F^{(l)}_{2k,k}$ are polynomials in Eisenstein series $E_2, E_4$ and $E_6$
\be\label{eq:F def}
    F^{(l)}_{2k,k} = \sum_{2k_2 + 4k_4 + 6k_6 = 2k} a^{(l)}_{k_2 k_4 k_6} E_2^{k_2} E_4^{k_4} E_6^{k_6} \;,
\ee
where $a^{(l)}_{k_2 k_4 k_6}$ are coefficients that will depend on the central charge $c$. Since we have specified that $F^{(l)}_{2k,k}$ is a quasi-modular form of depth at most $k$ we may have $a^{(l)}_{k_2k_4k_6} = 0$ for $k_2\leq k$. 

While the thermal correlator \eqref{eq:I pnt thermal corr} is a depth $K-1$ quasi-modular form we may have terms in the sums \eqref{eq:even weight MLDO} and \eqref{eq:odd weight MLDO} that have a greater depth. As an example consider the one point function $\Tr\left( \hat I_{5} q^{L_0 - \frac{c}{24}}\right)$. We use the ansatz \eqref{eq:even weight MLDO} together with \eqref{eq:F def} to obtain
\be\label{eq:J5 ansatz}\begin{split}
    \Tr\left( \hat I_{5} q^{L_0 - \frac{c}{24}}\right) =& \sum_{l=0}^{1} \sum_{k=0}^{3 - 3l} F^{(l)}_{2k,k} D^{3 - 3l - k} \Tr\left(W_0^{2l} q^{L_0 - \frac{c}{24}}\right) \\
    =& \left( a^{(0)}_{000} D^3 + a^{(0)}_{100} E_2 D^2 + \left( a^{(0)}_{010} E_4 + a^{(0)}_{200} E_2^2 \right) D \right. \\
   + &\left. \left( a^{(0)}_{001} E_6 + a^{(0)}_{110} E_2 E_4 + a^{(0)}_{300} E_2^3 \right)\right) \Tr\left( q^{L_0 - \frac{c}{24}} \right) + a^{(1)}_{000} \Tr\left( W_0^2 q^{L_0 - \frac{c}{24}} \right) \;.
\end{split}\ee
The thermal correlator has depth 0, but we have two terms, $E_2 D^2 \Tr\left( q^{L_0 - \frac{c}{24}} \right)$ and\\ $\Tr\left( W_0^2 q^{L_0 - \frac{c}{24}} \right)$, that transform as depth 1 quasi-modular forms. Hence these two pieces can be present because the additional terms appearing in their modular transformation can cancel each other to leave us with a depth 0 modular form. This can in fact be used to derive the modular transformation of $\Tr\left( W_0^2 q^{L_0 - \frac{c}{24}} \right)$ as was done in \cite{Iles:2013jha}. We also have a depth 2 and depth 3 piece, $E_2^2 D \Tr\left( q^{L_0 - \frac{c}{24}} \right)$ and $E_2^3 \Tr\left( q^{L_0 - \frac{c}{24}} \right)$ respectively. Since there is only one of each of these pieces the additional terms in their modular transformations cannot cancel to give us a depth 0 correlator and hence we must have $a^{(0)}_{200} = a^{(0)}_{300} = 0$.

Let us now justify the ansatz \eqref{eq:even weight MLDO} and \eqref{eq:odd weight MLDO} for thermal correlators. In the $\CW_3$ case our charges are composed of terms of the form (see \eqref{eq:I5B} for an example)
\be
    W_{n_1} \dots W_{n_I} L_{m_1} \dots L_{m_J} \;,
\ee
where we take a sum over integers $n_1,\dots,n_I$ and $m_1,\dots,m_J$ such that $\sum_{i=1}^I n_i + \sum_{i=1}^J m_i = 0$ and the product of modes will have a function of the indices in front of it. In the thermal correlator we therefore need to compute traces of the form
\be\label{eq:WL tr 0}
    \Tr\left( W_{n_1} \dots W_{n_I} L_{m_1} \dots L_{m_J} q^{L_0 - \frac{c}{24}} \right) \;.
\ee
We can commute any $L_0$ modes in the product to be next to the $q^{L_0-\frac{c}{24}}$ in the trace. In doing this we use the commutation relation \eqref{eq:L comm} which just gives an over all factor in front of the trace since
\be
    [L_0,L_{m_j}\dots L_{m_J}] = -(m_j + \dots + m_J)L_{m_j}\dots L_{m_J} \;.
\ee
We then replace the $L_0$ mode with a derivative since
\be\label{eq:L0 to div}
    L_0 q^{L_0-\frac{c}{24}} = \left( q\partial_q + \frac{c}{24} \right) q^{L_0-\frac{c}{24}} \;.
\ee
Hence we can replace any traces with $L_0$ modes in them with derivatives acting on traces of the form
\be\label{eq:WL tr}
    \Tr\left( W_{n_1} \dots W_{n_I} L_{m_1} \dots L_{m_J} q^{L_0 - \frac{c}{24}} \right) \;,
\ee
where $m_j \neq 0$ for all $j$. The process of replacing $L_0$ with a derivative reduces the number of $L_m$ modes in the trace. Let us now compute the trace \eqref{eq:WL tr}. We assume that $J\geq 1$. If this is not the case then we have only $W_{n}$ modes in our trace and we explain how to proceed below \eqref{eq:tr W only}. We first note the relation
\be
    L_{m_J} q^{L_0 - \frac{c}{24}} = q^{m_J} q^{L_0 - \frac{c}{24}} L_{m_J} \;.
\ee
Recalling that $m_J \neq 0$, we can use the above identity and the cyclicity of the trace to find
\be\begin{split}
    &(q^{-m_J} - 1)\Tr\left( W_{n_1} \dots W_{n_I} L_{m_1} \dots L_{m_J} q^{L_0 - \frac{c}{24}} \right) \\
    &= \sum_{i=1}^I \Tr\left( W_{n_1} \dots [L_{m_J}, W_{n_i}] \dots W_{n_I} L_{m_1} \dots L_{m_{J-1}} q^{L_0-\frac{c}{24}}\right) \\
    &+ \sum_{j=1}^{J-1} \Tr\left( W_{n_1} \dots W_{n_I} L_{m_1} \dots [L_{m_J}, L_{m_j}] \dots L_{m_{J-1}} q^{L_0-\frac{c}{24}}\right) \;.
\end{split}\ee
We can then evaluate the commutators using \eqref{eq:L comm} and \eqref{eq:LW comm}. During the evaluation of the commutator $[L_{m_J}, W_{n_i}]$, we keep the number of $W_n$ modes in the trace the same and reduce the number of $L_m$ modes by one. When we evaluate the $[L_{m_J}, L_{m_j}]$ commutator we produce two traces, one where the number of $L_m$ modes has been reduce by one and the other where the number of modes has been reduced by two. Hence this process replaces the trace \eqref{eq:WL tr} with several traces that all have the same number of $W_n$ modes as \eqref{eq:WL tr} and a reduced number of $L_m$ modes. If any of the new $L_m$ modes that appear from the commutation relations are $L_0$ then we can again replace them with a derivative, again reducing the number of $L_m$ modes. By iterating this process we will end up with commutators with $I$ modes $W_n$ and no $L_m$ modes.

We now have traces of the form 
\be\label{eq:tr W only}
    \Tr\left( W_{n_1} \dots W_{n_I} q^{L_0 - \frac{c}{24}} \right) \;,
\ee
to evaluate. We will assume that $n_I \neq 0$. If $n_I = 0$ then we can use the fact that $W_0$ and $L_0$ commute and the cyclicity of the trace to move the $W_0$ to the left hand side of the string of $W_n$ modes. Using the relation
\be
    W_{n_I} q^{L_0 - \frac{c}{24}} = q^{n_I} q^{L_0 - \frac{c}{24}} W_{n_I} \;,
\ee
we have
\be
    (q^{-n_I} - 1) \Tr\left( W_{n_1} \dots W_{n_I} q^{L_0 - \frac{c}{24}} \right) = \sum_{i=1}^{I-1} \Tr\left( W_{n_1} \dots [W_{n_I},W_{n_i}] \dots W_{n_{I-1}} q^{L_0 - \frac{c}{24}} \right)\;.
\ee
We can evaluate the commutation relations using \eqref{eq:W comm}. In doing so we reduce the number of $W_n$ modes by two and introduce three new trace terms. One with two $L_m$ modes, one with one $L_m$ mode and one with no $L_m$ modes. We can then commute these $L_m$ modes around the trace as we did above to remove them and be left with a collection of traces with now $I-2$ modes $W_n$. We can keep iterating this process to reduce the number of modes in the traces being evaluated. This algorithm terminates when we have a trace that only contains $W_0$ modes. We cannot simplify a trace of the form $\Tr(W_0^n q^{L_0-\frac{c}{24}})$ using this algorithm. Hence we have shown that every trace of the form \eqref{eq:WL tr 0} can be expressed as a collection of traces of the form
\be
   f(q) (q\partial_q)^m \Tr\left(W_0^l q^{L_0-\frac{c}{24}}\right) \;,
\ee
where $f(q)$ is a function of $q, c$ and the indices $n_i,m_j$ that appear in \eqref{eq:WL tr 0}. In the definition of the thermal correlators we sum over the indices and hence have the final form for a thermal correlator
\be\label{eq:diff op on tr}
    \Tr\left(\sum_{\sigma\in S_K} \hat I_{\sigma(n_1-1;a_1)} \dots \hat I_{\sigma(n_K-1;a_K)} q^{L_0 - \frac{c}{24}}\right) = \sum_{m,l} f_{m,l}(q) (q\partial_q)^m \Tr\left(W_0^l q^{L_0-\frac{c}{24}}\right) \;,
\ee
where $f_{m,l}(q)$ are functions of $q$ and $c$ and will depend on the specific thermal correlators in the trace. The sums over $m$ and $l$ are finite since we always start with a finite number of modes in the trace. 

We have shown that the thermal correlators of charges can be written as differential operators acting on $\Tr\left(W_0^n q^{L_0-\frac{c}{24}}\right)$. Since we are assuming that these thermal correlation functions are quasi-modular forms we then assume that these differential operators are modular linear differential operators. Hence we arrive at the ansatz for the thermal correlation functions given in \eqref{eq:even weight MLDO} and \eqref{eq:odd weight MLDO}.

It is complicated and lengthy to compute the thermal correlators by explicitly doing the required commutators explained above. Hence to compute the thermal correlators we instead start with the ansatz \eqref{eq:even weight MLDO} and \eqref{eq:odd weight MLDO} we can then explicitly compute the MLDO for a given thermal correlator by choosing a specific representation to trace over. The MLDO is universal so computing it in one theory gives it for all theories.

We will use the Verma modules to compute the MLDOs in the next section.

\section{Verma modules: traces and modular properties}
\label{sec:Verma}

Verma modules are especially simple modules of the $\CW_3$ algebra which are also irreducible for generic values of $c$, $h$ and $w$. This means that many (but not all) algebraic relations for Verma modules extend to generic representations. In this section we will use Verma modules to:
  express the particular traces \eqref{eq:TrW0n} for Verma modules as polynomials multiplying the Verma module character
  (section \ref{ssec:Vermatraces});
find the modular linear differential operators (MLDOs) discussed in the previous section (section \ref{ssec:Vermamldo});
 use the polynomial expression to find the modular properties of Verma module representations
(section \ref{sec:vermamodular});
 use the Verma module results to find expressions for the modular properties of the traces \eqref{eq:TrW0n} valid for all representations
(section \ref{sec:vermamodular2}).
We can then use these results in section \ref{sec:check} as evidence for our main conjecture.

\subsection{Verma module traces as polynomials}
\label{ssec:Vermatraces}

We now state the conjecture that when the representation $\CH$ is a Verma module, the traces \eqref{eq:TrW0n}  can be expressed as a polynomial in $c$, $\tilde h = h - \frac{c-2}{24}$, $w$ and the Eisenstein series multiplying the character of the Verma module. This was shown explicitly in \cite{Iles:2013jha} for $\Tr(W_0 q^{\Ione})$ and $\Tr(W_0^2 q^{\Ione})$ but the methods used are very involved and the method proposed here is much simpler (although relies on the conjecture being true)

Let us start with the thermal correlator $\Tr(W_0 q^{L_0 - \frac{c}{24}})$. It was shown in \cite{Iles:2013jha} that when we trace over a Verma module the thermal correlator takes the form
\be
    \Tr(W_0 q^{L_0 - \frac{c}{24}}) = w \Tr(q^{L_0 - \frac{c}{24}}) 
    = w \chi(q)\;,
    \label{eq:TrW1}
\ee
where 
\begin{align}
    \chi(q) = \Tr(q^{\Ione})
    = \frac{q^{h-(c-2)/24}}{\eta(q)^2}
    = \frac{q^{\tilde h}}{\eta(q)^2} \;,
\label{eq:Verma character}    
\end{align}
is the character of the Verma module representation,
$\eta$ is defined in \eqref{eq:eta derivative}
and we have defined 
$\tilde h = h-(c-2)/24$ for convenience.

From \cite{Dijkgraaf:1996iy}, we expect \eqref{eq:TrW1} to transform as a weight $3$ vector valued modular form but the character transforms as a weight $0$ (vector valued) modular form. 
This results suggests that we should treat $w$ as a graded object with grade 3

We can also consider the thermal correlator
\be\label{eq:SD to th}
    \Tr\left(\left(L_0 - \frac{c}{24}\right) q^{L_0 - \frac{c}{24}} \right) = \SD{0} \Tr( q^{L_0 - \frac{c}{24}} ) = \left(\tilde h -\frac{1}{12}E_2 \right) \chi(q) \;,
\ee
where $\tilde h = h - \frac{c-2}{24}$ and we have used 
\eqref{eq:eta derivative} to obtain the final expression. This transforms as a weight 2 modular form. This result suggests that we should treat $\tilde h$ as a graded object with grade 2.

Using the above results leads to the following conjecture for $\Tr(W_0^n q^{L_0 - \frac{c}{24}})$ when we trace over a Verma module. We already have the graded objects $w$ and $\tilde h$ with gradings 3 ans 2 respectively. We also treat the Eisenstein series $E_n$ as a graded object with grade $n$. We conjecture that the thermal correlators $\Tr(W_0^{n} q^{L_0 - \frac{c}{24}})$ can be written as a polynomial in $w,\tilde h$ and the Eisenstein series where the total grading of each term is $3n$,
\begin{align}
    \Tr(W_0^n q^{\Ione})
    = p_n(c,\tilde h,w,E_2,E_4,E_6)\,\chi(q)\;,
    \label{eq:pform}
\end{align}
where 
\be
    p_n  = \sum_{2k_{\tilde h}+3k_w+2k_2+4k_4+6k_6=3n} a^{(n)}_{k_{\tilde h} k_w k_2 k_4 k_6} w^{k_w} \tilde h^{k_{\tilde h}} E_2^{k_2} E_4^{k_4} E_6^{k_6} 
     \;,
     \label{eq:pansatz}
\ee
and the coefficients $a^{(n)}$ are polynomial's in $c$.
While we cannot justify this conjecture by commuting modes around the traces as we did in section \eqref{sec:MLDO generic reps}, we can test it by explicitly calculating these expressions and seeing that they agree with $\Tr(W_0^{n} q^{L_0 - \frac{c}{24}})$ to high order in $q$. 

Let us demonstrate the conjecture. The ansatz is trivial for $\Tr(W_0 q^{L_0 - \frac{c}{24}})$, it is
\be
    \Tr(W_0 q^{L_0 - \frac{c}{24}}) = a^{(1)}_{01000} w \Tr(q^{L_0 - \frac{c}{24}}) \;,
\ee
so there is just one coefficient to fix. If we just compute $\Tr(W_0 q^{L_0 - \frac{c}{24}})$ to leading order in $q$ we find
\be
    \Tr(W_0 q^{L_0 - \frac{c}{24}}) = q^{h-\frac{c}{24}}(w + O(q)) \;,
\ee
and therefore $a^{(1)}_{01000} = 1$ and we have
\be
    \Tr(W_0 q^{L_0 - \frac{c}{24}}) = w \Tr(q^{L_0 - \frac{c}{24}}) \;.
\ee
This agrees with the results from \cite{Iles:2013jha}. We now test the conjecture on a more non-trivial example. Let us compute $\Tr(W_0^2 q^{L_0 - \frac{c}{24}})$. From \eqref{eq:pansatz} we have
\begin{align}\label{eq:W02 ansatz}
    \Tr(W_0^2 q^{L_0 - \frac{c}{24}}) &=
    p_2 \chi(q)\;,
    \nonumber
    \\
    p_2 &=a^{(2)}_{02000} w^2 {+} a^{(2)}_{30000}\tilde h^3 {+} a^{(2)}_{20100}E_2 \tilde h^2 {+} (a^{(2)}_{10010}E_4 {+} a^{(2)}_{10200} E_2^2)\tilde h\nonumber\\ &\qquad {+} (a^{(2)}_{00001} E_6 \;{+} a^{(2)}_{00110}E_2E_4 {+} a^{(2)}_{00300}E_2^3) \;.
\end{align}
If we expand this as a power series in $q$, and remove the extraneous powers of $q$, we obtain
\begin{align}
    \Tr(W_0^2 q^{L_0 - h}) &= \Big(
a^{(2)}_{00001}+a^{(2)}_{00110}+a^{(2)}_{00300}+a^{(2)}_{02000} w^2+\tilde h (a^{(2)}_{10010}\nonumber\\&
\qquad
+a^{(2)}_{10200})+a^{(2)}_{20100} \tilde h^2+a^{(2)}_{30000} \tilde h^3\Big)
\nonumber\\
    &-2 \Big(251 a^{(2)}_{00001}-109 a^{(2)}_{00110}+35 a^{(2)}_{00300}-a^{(2)}_{02000} w^2
    \nonumber\\
    &\qquad -121 a^{(2)}_{10010} \tilde h+23 a^{(2)}_{10200} \tilde h+11 a^{(2)}_{20100} \tilde h^2-a^{(2)}_{30000} \tilde h^3\Big)q
    + O\left(q^2\right) \;.
\end{align}
We can then explicitly compute $\Tr(W_0^2 q^{L_0 - h})$ in the Verma module level by level and obtain
\be
    \Tr(W_0^2 q^{L_0 - h}) =  w^2 + \left( \frac{c-2}{36} + \frac{8}{3}\tilde h + 2w^2\right)q + O(q^2)  \;.
\ee
These two expansions should match for all $\tilde h,w$ and $q$. This gives us a system of over determined linear equations for the $a^{(2)}$ coefficients which we can solve to find
\be\begin{split}
    &a^{(2)}_{02000} = 1 \;,\; a^{(2)}_{30000} = a^{(2)}_{20100} = 0 \;,\; a^{(2)}_{10010} = - a^{(2)}_{10200} = \frac{1}{108} \;,\; a^{(2)}_{00001} = -\frac{3c+10}{77760} \;,\\
    & a^{(2)}_{00110} = \frac{c-10}{25920} \;,\; a^{(2)}_{00300} = \frac{1}{1944} \;.
\end{split}\ee
The final form for the thermal correlator is
\be
    \Tr(W_0^2 q^{L_0 - \frac{c}{24}}) = \left(w^2 {+} \frac{1}{108}(E_4 {-} E_2^2)\tilde h {+} \left(-\frac{3c {+} 10}{77760} E_6 {+} \frac{c {+} 10}{25920} E_2E_4 {+} \frac{1}{1944} E_2^3 \right) \right) \chi(q) \;.
    \label{eq:TrW02}
\ee
After some manipulation of the expressions it can be shown this matches the expression found in \cite{Iles:2013jha}.

We have two comments on the result \eqref{eq:TrW02}. First, although this is the result of the ansatz \eqref{eq:pansatz}, we note that the two terms in $\tilde h^3$ and $\tilde h^2$ vanish, and it is an empirical result that the sum over $k_{\tilde h}$ can be restricted to $k_{\tilde h} \leq n/2$.

Secondly, this is not the simplest form. Using the identities \eqref{eq:Enderivs}, it can be written as 
\be
    \Tr(W_0^2 q^{L_0 - \frac{c}{24}}) = \left(w^2 {-} \frac1 9 \tilde h  E_2'
    + \frac{1}{162
    } E_2 E_2'
    + \frac{3c + 10}{25920} E_4' 
    \right) \chi(q) \;.
    \label{eq:TrW02b}
\ee
It is another empirical result that the traces can always be written in a form with at least one derivative of an Eisenstein series in each term.

We have used this ansatz to calculate $\Tr(W_0^n q^{L_0 - \frac{c}{24}})$ up to 
$n=8$.
The results quickly become unmanageably large, we just give here the results for $n=3$ and $n=4$ making use of the identities \eqref{eq:Enderivs} to simplify them somewhat, 
\begin{align}
    \Tr(W_0^3 q^{L_0 - \frac{c}{24}}) &= \left(w^3 + 
    w\left( -\frac 13 \tilde h E_2' + \frac{1}{54}E_2 E_2' + \frac{3c+10}{8640}E_4'\right)
    \right) \chi(q) \;.
    \label{eq:TrW03}
\\    
    \Tr(W_0^4 q^{L_0 - \frac{c}{24}}) &= \left(w^4 + 
    w^2\left(- \frac 23 \tilde h E_2' + \frac{1}{27}E_2 E_2' + \frac{3c + 10}{4320}E_4'  \right)\right.
    \nonumber\\
    &\left. + \tilde h^2 \left(\frac 1 {36} E_4 E_2' - \frac{1}{324}E_2^2 E_2' - \frac 2{243} E_6'  \right)\right.
    \nonumber\\
    &\left. + \tilde h \left(
    \frac{9c-250}{38880}E_4E_2E_2'
    -\frac{87c - 670}{116640}E_6E_2' 
    +\frac{c-2}{3888}E_4 E_4'
    + \frac{5}{2916}E_2^3 E_2'
    \right) \right.
    \nonumber\\
    &+ \left(
    \frac{52100 + 18144 c - 945 c^2}{529079040}E_6 E_2 E_2' + 
    \frac{-333620 - 421740 c - 15741 c^2}{66512793600} E_4 E_4'\right.
    \nonumber\\
    &+ 
    \frac{-91060 + 420 c + 6363 c^2}{4702924800}E_4 E_2 E_4'
    + 
    \frac{-335884 - 8316 c - 2079 c^2}{13967686656} E_4 E_6'
    \nonumber\\
    &+\left.\left.
    \frac{310 - 63 c}{3149280}E_4 E_2^2 E_2'
    \right) \right) \chi(q) \;.
    \label{eq:TrW04}
\end{align}
To find the last expression it is necessary to calculate the traces in the Verma module up to order $q^4$. Note also that the highest power of $\tilde h$ is only $\tilde h^2$. One can find representations of this last expression with fewer terms by allowing $\chi'$ and $\chi''$, but so far we do not have a uniquely determined canonical form other than \eqref{eq:pform} which only involves $\chi$.

\subsection{Using Verma modules to find MLDOs}
\label{ssec:Vermamldo}

We now show how we can use Verma modules (i.e. specific representations) to find the MLDOs of section \ref{sec:MLDO generic reps} for any representation. The idea is to calculate the desired expression explicitly in the Verma module as a polynomial in $w$ and $\tilde h$ up to some level, and then use the corresponding results for $D^n \Tr(W_0^m q^{\Ione})$ to rewrite the expression in terms of $D^n \Tr(W_0^m q^{\Ione})$ and Eisenstein series. 

Let us demonstrate this technique with three examples. We will compute the thermal correlator for the Boussinesq charge $\hat I^B_4$ (or equivalently $[\hat W^3]_0$), and then for the product $W_0\,[\hat W^2]_0$. and finally for the two independent products of two $W_0$ and two $[\hat W^2]_0$ charges.

\subsubsection{Calculating \texorpdfstring{$\Tr(\hat I^B_4\,q^{\Ione})$}{Tr(I4B qI1)}}
The first example is the thermal correlator of the charge $\hat I^B_4$
. This is the image of the integral on the torus of the current $J_4^B$ (see appendix \ref{app:charges})
\begin{align}\label{eq:I5B}
    \int \! J_5^B\,\frac{\rd x}{2\pi}
    \mapsto 
    \hat I_4^\text{B} = 
    &
    \sum_{k> 0} (L_{-k}W_k + W_{-k}L_k)
    + W_0 L_0
    -\frac{c+6}{24} W_0\;.
\end{align}
In the Verma module we can compute the matrix $(\hat I_4^\text{B})_N$ which is the matrix of $\hat I_4^\text{B}$ in the level $N$ subspace of the Verma module. In the level $N$ subspace in which $L_0  = h + N $. We compute the trace of each $(\hat I_4^\text{B})_N$ to obtain
\begin{align}
    \Tr(\hat I_4^\text{B} q^{L_0-\frac{c}{24}}) 
    &= q^{h-\frac{c}{24}} 
    \left(
    w\Big(h - \frac{c{+}6}{24}\Big) 
    +
    2w\Big(h - \frac{c{-}90}{24}\Big) q
    +
    5w\Big(h - \frac{c{-}186}{24}\Big) q^2
    + O(q^3) 
    \right) \;.
    \end{align}
We can now compare this to the ansatz \eqref{eq:odd weight MLDO}, which takes the form here
\begin{align}
    \Tr(\hat I_4^\text{B} q^{L_0-\frac{c}{24}}) 
    &= 
    \Big(
    a^{(1)}_{000} \SD{0} + a^{(1)}_{100} E_2  
    \Big) \Tr(W_0\,q^{\Ione})\;.
\end{align}
We first have to calculate $\Tr(q^{L_0-\frac{c}{24}})$ and $\Tr(W_0 q^{L_0 - \frac{c}{24}})$ in the Verma module. Up to $O(q^3)$ we have
\begin{align}
    &\Tr(q^{L_0-\frac{c}{24}}) = q^{h-\frac{c}{24}}(1 + 2q + 5q^2 + O(q^3)) \;,\\
    &\Tr(W_0\, q^{L_0 - \frac{c}{24}}) = q^{h-\frac{c}{24}} w(1 + 2q + 5 q^2 + O(q^3))\;.
\end{align}
Full expressions to all orders in $q$ are known for these thermal correlators however they are not needed here. We now expand the ansatz as a power series in $q$ to obtain
\begin{align}
    \Tr(\hat I^B_4\,q^{L_0-\frac{c}{24}}) &= q^{h-\frac{c}{24}}\Big(
   \big(w( a^{(1)}_{100} 
   + a^{1}_{000}(h - \frac{c}{24})
   \big)
   +
   w\big(-22a^{(1)}_{100} 
   + a^{(1)}_{000}(2h+2-\frac{c}{12})
   \big)q
   \nonumber\\
   &\qquad
   +
   w\big(-115 a^{(1)}_{100}
   + a^{(1)}_{000}(5h+10-\frac{5c}{24})
   \big)q^2
    + O(q^3)\Big) \;.
\end{align}
These two expressions should match for all $h,w$ and $q$. Comparing the coefficients in the two expressions gives us a system of equations for the $a$ coefficients. This is an over determined set of linear equations which can be solved to find
\be
    a^{(1)}_{000} = 1 \;,\; a^{(1)}_{100} = -\frac{1}{4} \;.\; 
\ee
Hence the thermal correlator $\Tr(\hat I_4^\text{B} q^{L_0-\frac{c}{24}})$ is given by
\be\begin{split}
    &\Tr(\hat I_4^\text{B} q^{L_0-\frac{c}{24}}) =
     D^{(3)} \Tr(W_0\,q^{\Ione})\;,
\end{split}\ee
which we can write alternatively as 
\be\begin{split}
    &\Tr(\,[\hat W^3]_0\, q^{\Ione}) =
     8 \,\SD{3} \Tr(W_0\,q^{\Ione})\;,
     \label{eq:TrW3}
\end{split}\ee

\subsubsection{Calculation of traces of two charges}
\label{sec:2terms}

We can use the same method to calculate traces of products of charges. The first non-trivial example is 
\begin{align}
    \Tr(\,W_0\,[\hat W^2]_0\,q^{\Ione})
    = \frac23 \Tr(\,W_0\,\Lambda_0\,q^{\Ione})
    \;.
    \label{eq:TrW0Lambda0}
\end{align}
This is expected to be a weight 7 modular form, and so we make the ansatz
\eqref{eq:odd weight MLDO},
\begin{align}
     \Tr(W_0[\hat W^2]_0q^{\Ione})
     &=
     \Big(
     F^{(1)}_{0,0}\SD{5}\SD{3} + F^{(1)}_{2,1}\SD{3}
     + F^{(1)}_{4,2}\Big)\Tr(W_0\,q^{\Ione})
     \nonumber\\
     &= 
     \Big(
     a^{(1)}_{000} \SD{5}\SD{3}
     + a^{(1)}_{100} E_2 \SD{3}
     + a^{(1)}_{200}E_2^2 
     + a^{(1)}_{010}E_4 \Big)
     \Tr(W_0\,q^{\Ione})
     \;.
\end{align}
It is only necessary to calculate the traces explicitly up to order $q$ to fix the coefficients and obtain
\begin{align}
     \Tr(\,W_0\,[\hat W^2]_0\, q^{\Ione})
     &=
     \Big(
     \frac 23 \SD{5}\SD{3}
     + \frac13  E_2 \SD{3}
     +\frac{c{+}30}{2160}E_4 \Big)
     \Tr(W_0\,q^{\Ione})
     \;.
     \label{eq:TrW0W20}
\end{align}
which is a quasi-modular form of weight 7 and depth 1, as expected. We have checked explicitly that this is true up to order $q^8$.

\subsubsection{Calculation of traces of more charges}
\label{sec:4terms}

We can similarly calculate traces of larger numbers of charges by explicitly calculating the traces in the Verma module up to any particular order in $q$ and matching to a suitable ansatz. We must, however, take note of the order of the charges as we can only expect the traces to be quasi-modular forms of definite weight for symmetrised combinations.

It will be instructive to calculate the correlators of two non-commuting charges in which the order is important, and the first example which occurs in the expansion of $\exp(\a\t^3\hat\CW_0)$ are the correlators
\begin{align}
    f_{1212} \equiv 
    \Tr\Big(W_0\,[\hat W^2]_0\,W_0\,[\hat W^2]_0\,\,q^{\Ione}\Big)
    \;,\;\;
    f_{1122} \equiv 
    \Tr\Big((W_0)^2\,([\hat W^2]_0)^2\,q^{\Ione}\Big)
    \;.\;\;
    \end{align}
We expect from section \ref{sec:mldo} that the trace over the sum of all permutations of
$W_0$ and $[\hat W^2]_0$
 (using cyclicity of the trace),
\begin{align}
    2 f_{1212} + 4 f_{1122}  \;,
\end{align}
should be a quasi-modular form of weight 14.
We can also calculate the combination
\begin{align}
    (2f_{1212} + 4 f_{1122}) - 6 f_{1122}
    = \Tr\Big( \,
    \big[\,W_0\,,\,[\hat W^2]_0\,\big]
    \,
    \big[\,W_0\,,\,[\hat W^2]_0\,\big]
    \,\,q^{\Ione}\Big)
    \;.
\end{align}
We can calculate this commutator using the OPE 
\eqref{eq:WW2OPE},
\begin{align}
    \left[\,W_0\,,\,[\hat W^2]_0\,\right]
    = -\frac{8}{5}\,\hat J_0
    \;,\;\;\;\;\;\;
\end{align}
where 
$J$ is a quasi-primary field of weight 6 defined in equation \eqref{eq:Jdef}. This means that we expect
$\Tr((\hat J_0)^2 q^{\Ione})$ to be a quasi-modular form of weight 12.
This is indeed what we find - with the appropriate ansatz, and calculating the traces of $f_{1122}$ and $f_{1212}$ explicitly in the Verma module up to order $q^7$, we get
\begin{align}
    f_{1212} + 2f_{1122}&=
    \hbox{(quasi-modular, weight 14
    )}
    \;,
    \\
    f_{1212} - f_{1122} &= \hbox{(quasi-modular, weight 12
    )}
    \;.
\end{align}
The expressions themselves are rather lengthy and not especially instructive.
As expected, $f_{1212}$ and $f_{1122}$ are not quasi-modular forms of definite weight, but the appropriate combinations are.

We can similarly determine any other thermal correlators that appear. 


\subsection{Modular properties of Verma modules}
\label{sec:vermamodular}

It was already shown in \cite{Iles:2014gra} that one can use the free-field representation of Verma module representations to calculate the modular transforms of expressions of the form
\begin{align}
    p(w,\tilde h) \chi(q)\;.
\end{align}
We give the details in appendix \ref{app:Verma}, and just list a few results here. We will use the shorthand
\begin{align}
    p(w,\tilde h) \Tr(\hat q^{\Ione})
    \to p'(w,\tilde h) \Tr(q^{\Ione})
\end{align}
for     
\begin{align}
    p(w,\tilde h) \Tr_i(\hat q^{\Ione})
    = \sum_i S_{ij} p'(w,\tilde h) \Tr_j(q^{\Ione})
\end{align}
A list of some simple examples is given in \eqref{eq:transformexamples}, including in particular
\begin{align}
    w \chi(\hat q) & \to 
    \tau^3 w\; \chi(q)
        \;.\;\;
        \\
        w^2 \chi(\hat q) & \to 
    (\tau^6 w^2 - \frac{i}{3\pi}\tau^5\tilde h^2 - \frac{1}{3\pi^2}\tau^4\tilde h + \frac{i}{18 \pi^3}\tau^3 ) \chi(q)
        \;.\;\;
\end{align}

\subsection{Using Verma module properties to find results for all modules}
\label{sec:vermamodular2}

We are now in a position to calculate the modular transformation of the traces 
\eqref{eq:TrW0n}. We first 
calculate the corresponding polynomial in the Verma module, \eqref{eq:pform}
\begin{align}
    \Tr\left( W_0^n\, \hat q^{\Ione} \right) = p_n\,\chi(\hat q)\;.
\end{align}
We then use the method of section \ref{sec:vermamodular} to take the modular transform, using the results of appendix \ref{app:MF} for the transforms of the Eisenstein series.
Finally we use the results for the traces to express the polynomials in terms of traces and their derivatives.
We give this here for the case of $n=2$:
\begin{align}
    &\Tr(W_0^2 \hat q^{\Ione})
    \nonumber\\
    =& 
    \left(w^2 {-} \frac1 9 \tilde h  E_2'(-\tfrac{1}{\t})
    + \frac{1}{162
    } E_2(-\tfrac{1}{\t}) E_2'(-\tfrac{1}{\t})
    + \frac{3c + 10}{25920} E_4'(-\tfrac{1}{\t}) 
    \right) \chi(\hat q) \;.
\nonumber\\
 \to&
 \Big((\t^6 w^2 - \frac{i}{3\pi}\t^5\htilde^2 - \frac{1}{3\pi^2}\tau^4\htilde + \frac{i}{18 \pi^3}\t^3)
 -\frac{1}{9}
 (\t^2\htilde -\frac{i}{2\pi}\t)( \t^4E_2'(\t)  - \frac{i}{\pi}\t^3 E_2(\t)
 -\frac{3}{\pi^2}\t^2) 
   \nonumber\\
 & 
 +\frac{1}{162}
 (\t^2 E_2 - \frac {6i\t}{\pi})
 (\t^4E'_2(\t)  - \frac{i}{\pi}
\t^3 E_2(\t) -\frac{3}{\pi^2}\t^2 ) 
 + \frac{3c{+}10}{25920}(\tau^6 E_4'(\t)  - \frac{2i}{\pi}\tau^5 E_4(\t)
 \Big)\chi(q)
 \nonumber\\
 =&{}\;\Big(
 \t^6( w^2 - \frac 19 \htilde E_2'(\t) + \frac 1{162
 }E_2(\t)E_2'(\t) + \frac{3c+10}{25910}E_4'(\t)) 
 \nonumber\\
 &+\t^5 (
 -\frac{(3c{+}10)}{12960\pi}E_4(\t)
 + \frac{i}{162\pi}(3E_2'(\t)-E_2^2(\t))
 + \frac{i}{9\pi} \htilde E_2(\t)
 -\frac{i}{3\pi}\htilde^2
 )\Big)\chi(q)
 \nonumber\\
 =&{}\;
 \t^6 \Tr(W_0^2q^{\Ione}) 
 +\frac{\t^5}{3\pi i} \left( \SD 2 \SD 0 +\frac{ c E_4(\t)}{1440}\right) \Tr(q^{\Ione})
\;,
\end{align} 
in agreement with \eqref{eq:W0sq}.

The important fact to note is that this result is now independent of the special properties of the Verma module that were used in its derivation - it is in fact true for all representations of the $\CW_3$ algebra.

An advantage of this method is that it is straightforward to programme (eg in Mathematica) and we have used this to calculate the modular transforms of $\Tr(W_0^n q^{\Ione})$ up to $n=8$.


\section{Complete check of the conjecture at \texorpdfstring{$c=-2$}{c=-2}}
For general theories we have only been able to check our conjecture perturbatively (up to order $\a^7$), as mentioned at the end of the previous section. As of now we have no proof in the general case. However when we restrict to a special value of the central charge, namely $c = -2$, we can prove our conjecture. We outline this proof in the following section.

\label{sec:c=-2}
\subsection{The Symplectic Fermion}
The $\CW_3$ algebra at $c=-2$ can be realised by the free field theory known as the Symplectic Fermion. It has a rich representation theory but we will focus only on the representations whose characters' modular properties are well behaved, that is the untwisted and half-twisted representations. For more details on the Symplectic Fermion, see \cite{Creutzig:2013hma,Kausch:2000fu}.

The operator algebra of the Symplectic Fermion is
\begin{equation}
    \chi^{\a}(z)\chi^{\b}(w) = \frac{\epsilon^{\a\b}}{(z-w)^2}, \quad \a,\b = \pm, \quad \epsilon^{+-} = -\epsilon^{-+} =1.
\end{equation}
This implies a mode algebra
\begin{equation}
    \{\chi^{\a}_m,\chi^\b_n\} = m\epsilon^{\a\b}\delta_{m+n,0} \;.
\end{equation}
These are Virasoro primary fields of weight $1$ at $c=-2$ with respect to the stress tensor
\begin{equation}
    T(z) = (\chi^-\chi^+)(z) \;.
\end{equation}
The representations that will be relevant for this work are the irreducible vacuum module $\bb{L}_0$ built out of the vacuum state $\ket{0}$; and the irreducible half-twisted module $\bb{L}_{1/2}$, built out of a primary state $\ket{-1/8}$ with weight $h=-1/8$. These modules are constructed as Fock spaces,
\begin{equation}\label{eq: L_0}
    \bb{L}_0 = \text{span} \left\{ \chi^-_{-n_1}\dots\chi^-_{-n_\nu}\chi^+_{-m_1}\dots\chi^+_{-m_\mu}\ket{0} \,\,\textbf{;} \,\,\begin{matrix}
        \mu\geq0,\\ \nu\geq0,
    \end{matrix}, \begin{matrix}
        n_i > n_{i+1}>0,\\m_{i}>m_{i+1}>0,
    \end{matrix}\,\, n_i,m_i \in \bb{Z}\right\},
\end{equation}
\small{
\begin{equation}\label{eq: L_1/2}
    \mathbb{L}_{1/2} = \text{span}\left\{     \chi^-_{-n_1}\dots\chi^-_{-n_\nu}\chi^+_{-m_1}\dots\chi^+_{-m_\mu}\ket{-\tfrac{1}{8}} \,\,\textbf{;} \,\,\begin{matrix}
        \mu\geq0,\\ \nu\geq0,
    \end{matrix}, \begin{matrix}
        n_i > n_{i+1}>0,\\m_{i}>m_{i+1}>0,
    \end{matrix} \,\, n_i,m_i \in \bb{Z}+\tfrac{1}{2}\right\}.
\end{equation}}
\normalsize
One can give these spaces a $\bb{Z}_2$ grading by introducing the fermion number operator $F$ which has the property:
\begin{equation}
    \{\chi^\a_{m},(-1)^F\}=0 \quad,\quad F \ket{0} =F \ket{-\tfrac{1}{8}} = 0.
\end{equation}
In an upcoming work \cite{KarimiWatts}, we construct an infinite set of bilinear fields in this theory which one can integrate to obtain an infinite set of mutually commuting conserved charges.
\begin{equation}\label{eq: bilinears}
\begin{split}
 B_k(w) &= (\del^{k-2}\chi^-\chi^+)(w) \mod \del \quad,\quad k \in \{2,3,4,...\},\\
        \Q_{n} &= \int_{0}^{2\pi i}\frac{dw}{2\pi i}B_{n+1}(w) \Rightarrow [\Q_m,\Q_n]=0 \;.\\
\end{split}
\end{equation}
The notation ``$\text{mod}\ \del$" denotes total derivative terms that fix the respective fields to be quasi-primary. Acting on the module $\bb{L}_{\lambda}$ the charge is
\begin{equation}\label{eq: THE CHARGES}
    \Q_{m} = (-1)^{m-1} \sum_{j\in \bb{Z}+\lambda} j^{m-1}:\chi^- _j \chi^+_{-j}: - \;{\bf c}_{m}(\lambda)  \quad,\quad \lambda = 0,\frac{1}{2} \;.
\end{equation}
In the above expression we have
\begin{equation}
    {\bf c}_{m}(\lambda) = \begin{cases}
         \zeta(-m)& \lambda =0\;,\\
         \left(\left(\tfrac{1}{2}\right)^{m}-1\right)\zeta(-m)& \lambda = 1/2\;.
    \end{cases}
\end{equation}
The commutation relations with the Symplectic Fermion modes follow directly from this
\begin{equation}
    [\Q_m,\chi^\pm_n] = \mp (-n) ^m \chi^\pm_n \;.
\end{equation}
These bilinear fields realise the $\CW_3$ algebra by the identification $T  = B_2$ and $W = \frac{1}{\sqrt{6}}B_3$. Define the following traces
\begin{equation}
    \Tr_{\lambda,{1/2}}(\SCO) = \Tr_{\bb{L}_\lambda}(\SCO),\quad \Tr_{\lambda,0}(\SCO) = \Tr_{\bb{L}_\lambda}((-1)^F\SCO) \;.
\end{equation}
Then the GGE we're interested in is
\begin{equation}
\chi_{\lambda,s}(\t,\a) = \Tr_{\lambda,s}(q^{L_0 - c/24}e^{\a \Q_2}) = q^{h(\lambda) + \frac{1}{12}}\prod_{n\in \bb{N} -\lambda}(1 - e^{2\pi i s} q^ne^{\a n^2})(1 - e^{2\pi i s} q^ne^{-\a n^2}) \;.
\end{equation}
Here we have that $h(0)=0$ and $h(1/2) = -1/8$. When $\a=0$, these are just the characters of the Symplectic Fermion which transform as
\begin{equation}\label{eq: S-transform characters}
\begin{split}
    \begin{pmatrix}
    \chi_{0,1/2}(-1/\t)\\\chi_{1/2,1/2}(-1/\t)\\\chi_{1/2,0}(-1/\t)\\ 
\end{pmatrix}  &=  \begin{pmatrix}
    0 &0&1/2\\0&1&0\\2&0&0
\end{pmatrix} \begin{pmatrix}
    \chi_{0,1/2}(\t)\\ \chi_{1/2,1/2}(\t)\\ \chi_{1/2,0}(\t)\\ 
\end{pmatrix},\\
\chi_{0,0}(-1/\t) &= - i \t\chi_{0,0}(\t) \;.\\
\end{split}
\end{equation}

\subsection{Asymptotic analysis of exact transformation}
In the upcoming work \cite{KarimiWatts}, we show that the exact transformation of the GGE, after defining $\a = 2\pi i \hat{\t}^2 \a_2$ and $\hat{\t}= -1/\t$, is as follows,
\begin{equation}\label{eq: the transformation exact}
\begin{split}
    \chi_{\lambda,s}(\hat{\t},\a) =\frac{2^{\delta_{s,0}\delta_{\lambda,1/2}}}{2^{\delta_{s,1/2}\delta_{\lambda,0} }}\left(\frac{\t}{i}\right)^{\delta_{\lambda,0}\delta_{s,0}}q^{h_0^{(s)}(\t,\a)} \prod_{\pm}\prod_{n \in \bb{Z}-s}\prod_{ z^{\pm}_j(n) \in \bb{H}}(1-e^{2\pi i \lambda}e^{2\pi i  z_j ^ \pm(n)}) \;,
\end{split}
\end{equation}
where
\begin{equation}
    h_0^{(s)}(\t,\a)=\frac{1}{2\pi i\t}\sum_{\pm} \biggl \{  \int_{0}^\infty (\log(1-e^{2\pi i s}e^{2\pi i (\hat{\t} x  \pm \hat{\t}^2x^2\a_2)})) dx\biggr \} \;,
\end{equation}
and $z_j^{\pm}$ is a solution to a polynomial equation\footnote{Note that here $\pm$ labels the equation, and that $j$ is an index for the solutions to that equation.}
\begin{equation}
\label{eq: Polynomial for transform}
    z = z_j^{\pm}(\omega): \qquad (\hat{\t} z) \pm \a_2 (\hat{\t} z)^2 = \omega \;.
\end{equation} 
Without loss of generality, we analyse the particular example of the transformation\\ $\chi_{1/2,0}(-1/\t) = 2 \chi_{0,1/2}(\t)$. However, the work done here is in essence the same for all the representations in the transformations in \eqref{eq: S-transform characters}. To recover the main conjecture of this paper, we need to take an asymptotic expansion about $\a \to 0$.
Only one solution from each of these polynomial equations is finite in that limit. The divergent terms in this limit will not contribute to the transformation, as can be seen from \eqref{eq: the transformation exact}, since they will either not belong to the upper half-plane or will vanish in the exponential.

The piece of the transformation which is non-trivial in the asymptotic regime is
\begin{equation}\label{eq: asym, transform, 1/2,0}
    2 \, q^{h_0^{(0)}(\t,\a)}  \prod_{n=1}^{\infty}(1+q^ne^{n f(n)})(1+q^{n}e^{nf(-n)}),
\end{equation}
where
\begin{equation}
f(n) =\sum_{k=0}^{\infty}\frac{i 2^{k+2} \pi ^{-k} \tau  n^{k+1} \left(\frac{1}{2}\right)_{k+1} \left(-i \alpha  \tau ^2\right)^{k+1}}{(2)_{k+1}},
\end{equation}
and
\begin{equation}
    h_0^{(0)}(\t,\a) = \frac{1}{\a \tau^2}\sum_{k=0}^{\infty} \frac{2^{-4k-2}(2\pi i)^{2k+2}}{(2k+1)} \binom{4k}{2k}\left(\frac{\a\t^2}{\pi^2}\right)^{2k+1}\zeta(-2k-1).
\end{equation}
After some massaging, one can show that  \eqref{eq: asym, transform, 1/2,0} is the expression of a GGE in the untwisted representation $\bb{L}_0$. Indeed, we find that
\begin{equation}
\chi_{1/2,0}(\hat{\t},\a) \sim 2 \Tr_{0,1/2}(q^{L_0-\frac{c}{24}}e^{\hat{C}}),
\end{equation}
where
\begin{equation}\label{eq: Charge from limit}
\hat{C} 
        = \a\t^3 \sum_{k=0}^{\infty}\left(\left(\frac{\a\t^2}{4\pi i}\right)^{k}\frac{1}{k!}\frac{2^{k+1}(2k+1)!}{(k+2)!}\Q_{k+2}\right).
\end{equation}
Repeating this process with all the representations and bringing them together, one finds what one expects, that is,
\begin{equation}
    \Tr_i(\hat{q}^{L_0-c/24}e^{\a\Q_2}) \sim\sum_jS_{ij}\Tr_j(q^{L_0-c/24}e^{\hat{C}}).
\end{equation}

\subsection{Comparison with our conjecture}
We would like to compare the asymptotic limit of the exact result \eqref{eq: Charge from limit} with our conjecture \eqref{eq:conj1a}. When applying this conjecture to the symplectic fermion, it is important to remember that the $\CW_3$ algebra is realised differently ($B_3 = \sqrt{6}W$). This is so that the field $W$ may be brought inline with the other bilinear fields present in the symplectic fermion \eqref{eq: bilinears}, as these are the fields that will appear in the modular transformation. All this means is that we have to take into account this normalisation when we construct the charge appearing in the modular transformation by replacing $W$ with $B_3$ in $[W^{k+1}]$ appearing in \eqref{eq:conj1a}. The conjecture \eqref{eq:conj1a} is
\begin{equation}
    \Tr_i(\hat{q}^{L_0-c/24}e^{\a\Q_2}) \sim \sum_jS_{ij}\Tr_{j}(q^{L_0-c/24}e^{\hat{\CB}})\;,
\end{equation}
where
\begin{equation}\label{eq: C hat Symp Ferm}
    \hat{\CB} = \a\t^3 \sum_{k=0}^{\infty}\left(\left(\frac{\a\t^2}{4\pi i}\right)^{k}\frac{1}{k!}[B_3^{k+1}]_0\right).
\end{equation}
Our task is then to show that $\hat{\CB}$ matches $\hat{C}$. What is effectively the same as in \cite{Dijkgraaf:1996iy} is that the bilinear fields satisfy a pre-lie algebra structure and the fields have simple second-order poles in their OPEs. We elaborate on this point in \cite{KarimiWatts}, but this result is not surprising. What we will need for our work is just the second-order pole:
\begin{equation}\label{eq: Pre-Lie}
    (B_mB_n\lb = (m+n-2)B_{m+n-2} \mod \del.
\end{equation}
Using the relation \eqref{eq: Pre-Lie}, and also relying on the recursion relation that is elaborated in Appendix \ref{app:otter},
\begin{equation}\label{eq: Conjectured recursion}
    [B_3^{k+1}] = \sum_{m=1}^k \binom{k-1}{m-1}([B_3^m][B_3^{k-m+1}]\lb,
\end{equation} 
we have proven that:
\begin{equation}\label{eq: defintion of a(k)}
    [B_3^k] =a(k)B_{k+2},\quad a(k):=\frac{2^k(2k-1)!}{(k+1)!}.
\end{equation}
This can be proven by noticing that the recursion relation \eqref{eq: Conjectured recursion} is equivalent to the generating function $f(x) = \sum_{k=0}^{\infty} \frac{a(k+1)}{k!}x^k$ satisfying an ODE that is solved by
\begin{equation}
        f(x) = \frac{-4x -\sqrt{1-8x}+1}{8x^2}.
\end{equation}
One can check that the coefficients of this function in a series expansion will give \eqref{eq: defintion of a(k)}, and one can also check numerically for low-lying cases that this is precisely what we expect from \eqref{eq: Pre-Lie}. This reduces the conjectured transformation to
\begin{equation}
        \hat{\CB} = \a\t^3 \sum_{k=0}^{\infty}\left((\tfrac{\a\t^2}{4\pi i})^{k}\tfrac{1}{k!}a(k+1)\Q_{k+2}\right).
\end{equation}
This is exactly the expression \eqref{eq: Charge from limit}. 

One remark worth making with regard to this check is that, due to the relation \eqref{eq: Pre-Lie}, we do not have to worry about the commutativity of the charges appearing in the modular transformed GGE. That is, due to that relation, we will always obtain zero-modes $\Q_n$ of bilinear fields $B_{n+1}$ in the transformed GGE. By construction, all of the charges $\Q_n$ commute with one another, so no issues with non-commutativity of the resultant charges could appear.

\section{Outlook}

 A two-dimensional conformal field theory (CFT) with a $\mathcal{W}_3$ symmetry algebra possesses infinitely many conserved charges, making it natural to define the partition function as a generalized Gibbs ensemble (GGE). In this work, we propose a conjecture for the modular $S$-transformation of the GGE in the presence of the simplest non-trivial charge, namely, the second quantum Boussinesq charge. We emphasize that this conjecture is an asymptotic result rather than an exact one. While a rigorous proof remains elusive, the conjecture has been extensively verified using multiple independent methods.
 
 Following the approach of \cite{Gaberdiel:2012yb}, we compute the modular $S$-transformation of the trace of zero modes for arbitrary fields. This requires the evaluation of thermal $n$-point functions using Zhu recursion relation. However, for the case of $n$ insertions of the $W$ field, the computation becomes challenging for $n \geq 5$. Consequently, we have restricted our analysis to the modular $S$-transformation of the asymptotic series up to fourth order in the chemical potential and found agreement with our proposed formula.

Additionally, we employ the methods of \cite{Iles:2014gra} to compute the modular $S$-transformation of the asymptotic series up to seventh order ($n \leq 7$) on a Verma module. This approach provides better computational control compared to the previous method, allowing us to extend the analysis to higher orders in the asymptotic expansion. As a further consistency check, we consider the case of $c\,{=}\,{-}2$ symplectic fermions, where the modular $S$-transformation can be computed to all orders in $n$, and find precise agreement with our conjecture.    

 Since our proposed modular transformation formula remains conjectural at this stage, representing an asymptotic result rather than an exact identity, establishing a rigorous proof 
 (for example by extending the analysis of \cite{Dijkgraaf:1996iy} to the case where the pre-Lie algebra condition is not met) constitutes an important open problem that we intend to address in future work, even though the current evidence strongly supports its validity. A particularly significant advancement would be the derivation of an exact closed-form expression, analogous to the complete solution available for the $c\,{=}\,{-}2$ symplectic fermion system.
 
 Furthermore, our preliminary analysis suggests that an analogous transformation law might exist for generalized Gibbs ensembles incorporating non-trivial KdV charges for which an extensive analysis is needed. 

 A very natural generalisation can be made for a GGE charged with multiple conserved charge with non-zero chemical potential.

 In the framework of holography, higher-spin theories in anti-de Sitter (AdS) spacetime incorporate gauge fields with spin $s>2$ alongside the graviton and are dual to conformal field theories (CFTs) with extended symmetry algebras. In the specific case of  $\text{AdS}_3$,  this duality connects higher-spin gravity theories to two-dimensional CFTs endowed with extended $\mathcal{W}$ algebras. A prominent example is the $\text{AdS}_3/\text{CFT}_2$  correspondence, where higher-spin gravity—formulated as a Chern-Simons theory with gauge group  $SL(3,\mathbb{R})\times SL(3,\mathbb{R})$ is holographically dual to a 2D CFT with an extended $\mathcal{W}_3$ algebra \cite{Campoleoni:2010zq,Henneaux:2010xg}. Building on this correspondence, \cite{Gutperle:2011kf} investigated three-dimensional higher-spin gravity in the Chern-Simons formulation with gauge group  $SL(3,\mathbb{R})\times SL(3,\mathbb{R})$, deriving black hole solutions carrying higher-spin charges and computing their entropy.  Since  our proposed conjecture gives us the modular transformation of the partition function, it enables the derivation of a Cardy-like formula, analogous to \cite{Hartman:2014oaa}, which reproduces the gravitational entropy in the large $c$ limit.  Moreover, since our results are exact in the central charge, they also permit the systematic computation of sub-leading corrections to the entropy in powers of $1/c$.

\section*{Acknowledgements}
%

FK thanks William H. Pannell 
and MD, TS and AS thank Sujay K. Ashok for helpful discussions related to this work, and we are all grateful to him for a critical reading of the manuscript. TS would also like to thank Ronak M. Soni for helpful discussions. AS thanks Amit Suthar for valuable discussions. GW thanks Senan Sekhon for discussions on Otter trees.

MD is supported by the French Agence Nationale de la Recherche (ANR) under grant ANR-21-CE40-0003 (project CONFICA).
FK is supported by an STFC studentship under grant  ST/Y509279/1.
GMTW acknowledges partial support from STFC grant  ST/T000759/1.

 For the purposes of open
access, the authors have applied a Creative Commons Attribution (CC BY) license to any
Accepted Author Manuscript version arising from this submission.

\appendix

\section{The \texorpdfstring{$\CW_3$}{\CW3}-algebra and its representation theory}

\label{sec:Wreps}

For the $\CW_3$ algebra we will use the conventions from \cite{Ashok:2024zmw,Ashok:2024ygp} where the operator product expansions of the fields
\begin{align}
    T(z) = \sum_m L_m z^{-m-2}
    \;,\;\;
    W(z) = \sum_m W_m z^{-m-3}
    \;,
\end{align}
are
\begin{align}
    T(z) T(w) &= 
    \frac{c/2}{(z-w)^4} + \frac{2 T(w)}{(z-w)^2} + \frac{T'(w)}{(z-w)} + O(1)
    \;,\;\;
    \\
T(z) W(w) &= 
    \frac{3 W(w)}{(z-w)^2} + \frac{W'(w)}{(z-w)} + O(1)
    \;,\;\;
    \\
    W(z) W(w) &= 
    \frac{c/(9 b^2)}{(z-w)^6} 
    + \frac{1}{3b^2}\left(\frac{2 T(w)}{(z-w)^4} + \frac{T'(w)}{(z-w)^3} + 
    \frac 3{10}\frac{T''(w)}{(z-w)^2} + \frac 1{15}\frac{T'''(w)}{(z-w)}\right)
    \nonumber\\&\qquad
    + \frac23\frac{\Lambda(w)}{(z-w)^2}
    + \frac13\frac{\Lambda'(w)}{(z-w)} + O(1)
    \;,
\label{eq:WWope}    
\end{align}
where $b^2 = \frac{16}{22+5c}$. 
Here $\Lambda$ is the quasi-primary field of weight 4 corresponding to the state
\begin{align}
    \ket\Lambda = \Big(L_{-2}L_{-2} - \frac 35 L_{-4} \Big)\vac
\end{align}
with mode expansion
\begin{align}
    \Lambda(z) = \sum_m \Lambda_m\, z^{-m-4}\;.
\end{align}
The modes $\Lambda_n$ are
\be
    \Lambda_n = \sum_{k\in\Z} :L_k L_{n-k}: + \frac{1}{5}x_n L_n \;,
\ee
where $x_{2l} = 1 - l^2$ and $x_{2l+1} = 2 - l - l^2$ and the normal ordering $:L_k L_{n-k}:$ means we put the operator with larger index on the right.

The commutators of the modes of the $\CW_3$ algebra are correspondingly
\begin{align}
    [L_n,L_m] =& (n-m)L_{n+m} + \frac{c}{12}n(n^2-1) \delta_{n+m,0} \;, \label{eq:L comm}\\
    [L_n,W_m] =& (2n-m) W_{n+m} \;,\label{eq:LW comm}\\
    [W_n,W_m] =& \frac{n-m}{3}\Lambda_{n+m} + \frac{n-m}{3b^2}\left(\frac{(n+m+3)(n+m+2)}{15} - \frac{(n+2)(m+2)}{6}\right)L_{n+m} \nonumber\\
    &+ \frac{c}{1080b^2}n(n^2 - 4)(n^2 - 1)\delta_{n+m,0} \label{eq:W comm}\;,
\end{align}

The Verma modules of the $\CW_3$ algebra are labelled by the central charge $c$, and two complex numbers $h$ and $w$. The Verma module is spanned by the states
\begin{align}
    W_{i_1}..W_{i_m} L_{j_1}.. L_{j_n} | h,w\rangle
    \;,\;\;\;\;
    i_k \leq i_{k+1} <0
    \;,\;\;
    j_k \leq j_{k+1} <0    
\;,
\end{align}
where $|h,w\rangle$ satisfies
\begin{align}
    L_m |h,w\rangle = W_m |h,w\rangle = 0\;, m>0\;\;\;
    L_0|h,w\rangle = h |h,w\rangle
    \;,\;\;
    W_0 |h,w\rangle = w |h,w\rangle\;.
\end{align}

\section{Charges of integrable hierarchies}
\label{app:charges}

We list here the first few currents of the KdV and Boussinesq hierarchies (in our normalisations), both on the cylinder ($J^{\text{KdV/B}}_n$) and after the map to the plane
($\hat J^{\text{KdV/B}}_n)$ (in the notation of appendix \ref{app:Gaberdiel}). To avoid ambiguity, we give these in terms of the corresponding state through the state-field correspondence. There is a non-trivial KdV current  $J^{\text{KdV}}_n$ for all even $n$, and a Boussinesq current $J^B_n$ for all $n\equiv 0,2$ mod 3. The currents $J$ are only defined up to total derivatives; here we give the representative that is quasi primary, even though this is not the simplest form. 
\begin{align}
 \ket{J^B_3 }&= W_{-3}\vac\;,\;\;
 \\
 \ket{\hat J^B_3} &= \ket{J^B_3}\;,\;\;
 \\
 \ket{J^{\text{KdV}}_4}&= \left(\,L_{-2}L_{-2} - \frac 35 L_{-4}\,\right)\vac
 \;,\;\;
\\
 \ket{\hat J^{\text{KdV}}_4}&= \ket{J^{\text{KdV}}_4} + 
 \left(\,- \frac{5c{+}22}{60} L_{-2} + \frac{c(5c{+}22)}{2880}\,\right)\vac
 \;,\;\;
 \\
 \ket{J^B_5} &=
 \left(\,
 W_{-3}L_{-2} - \frac{10}{7}W_{-5}
 \,\right)\vac
\;,\;\;
 \\
 \ket{\hat J^B_5} &=
 \ket{J^B_6} + 
 \left(\,
 -\frac{7c{+}114}{168}W_{-3}
 \,\right)\vac
 \\
 \ket{J^{\text{KdV}}_6} &=
 \left(\,
 L_{-2}L_{-2}L_{-2}
 + \frac{5c{-}98}{189} L_{-6}
 + \frac{2c{-}23}{27}L_{-4}L_{-2}
 - \frac{5c{+}64}{108}L_{-3}L_{-3}
 \,\right) \vac     
 \;,\;\;
 \\
 \ket{\hat J^{\text{KdV}}_6} &=
 \ket{J^{\text{KdV}}_6} +
 \left(\,
 -\frac{5(7c{+}68)}{216} L_{-2}L_{-2}
 +\frac{1904 {+} 128 c {-} 7 c^2}{2268} L_{-4} 
 - \frac{5 (476 {+} 185 c {+} 14 c^2) }{9072}L_{-3}\right.
 \nonumber\\
 &\qquad\left.
+ \frac{952 {-} 242 c {-} 35 c^2}{36288} L_{-2}
- \frac{c (952 {+} 302 c {+} 21 c^2) }{290304}
 \,\right) \vac
 \;,
 \\
 \ket{J^{\text{B}}_6} &=
 \left(
 9 W_{-3}W_{-3} {+} L_{-2}L_{-2}L_{-2}
  -\frac{65c{-}206}{252}L_{-6}
  +\frac{c{-}166}{36}L_{-4}L_{-2}
  \right.\nonumber\\&\qquad\left.
  - \frac{5c{+}574}{288}L_{-3}L_{-3}
 \right) \vac
 \;,\;\;
 \\
 \ket{\hat J^{\text{B}}_6} &=
 \ket{J^{\text{B}}_6} +
 \left(\,
 -\frac{5(c{+}23)}{36} L_{-2}L_{-2}
 +
 \frac{10718 {+} 305 c {-} 7 c^2}{6048} L_{-4}
 - \frac{5 (874 {+} 199 c {+} 7 c^2)}{12096}L_{-3}
 \right.
 \nonumber\\
 &\qquad \left.
 + \frac{2714 {+} 923 c {+} 
    35 c^2}{12096} L_{-2}
    -\frac{c (690 {+} 191 c {+} 7 c^2)}{96768}
 \,\right) \vac
 \;.\;\;
\end{align}
The charges (after the map to the plane) are given by the zero modes of the corresponding currents:
\begin{align}
    \hat I_{n-1}^{\text{KdV/B}} = (\hat J^{\text{KdV/B}}_{n})_0 \;.
\end{align}
The currents $J_n$ are defined only up to total derivatives, and the charges $\hat I_{n-1}$ are independent of any total derivative terms in $J_n$.

\section{Properties of \texorpdfstring{$[W^m]$}{[Wm]}}
\label{app:otter}

We have defined the field $[W^k]$ through the recursion relation \eqref{eq:conj1b} leading to the following expressions:
\begin{align}
    [W] &= W
    \;,\\
    [W^2] &= (WW\lb
    \;,\\
    [W^3] &= ((WW\lb W\lb + (W(WW\lb \lb
    \;,\\
    [W^4] &= (((WW\lb W\lb W\lb  + ((W(WW\lb \lb W\lb 
    + ((WW\lb (WW\lb \lb 
    \nonumber\\ &+ ((WW\lb (WW\lb \lb + (W((WW\lb W\lb \lb  + (W(W(WW\lb \lb \lb
    \;,\\
    &\vdots
\end{align}
It is a fundamental property of the bracket $(AB\lb $ that it is commutative (up to total derivatives), and so one can reduce these expressions by moving brackets to the right, to get (where, here only, we abbreviate 
$(XY)\lb \equiv(XY)$,)
\begin{align}
    [W] &= W
    \;,\\
    [W^2] &= (WW)
    \;,\\
    [W^3] &= 2(W(WW))
    \;,\\
    [W^4] &= 2 ((WW)(WW))  + 4 (W(W(WW(((
    \;,\\
    [W^5]
    &= 8 (W(W(W(WW)))) + 12((WW)(W(WW))) + 4 (W((WW)(WW)))   \\[1mm]
    [W^6] 
    &= 16(W(W(W(W(WW))))) + 
    \ldots 
\;.
\end{align}
We also give here the states corresponding to the first few of these fields,
\begin{align}
    \ket{\,[W]\,} &= W_{-3} \vac
    \;,\;\;
    \nonumber\\
    \ket{\,[W^2]\,} &= \Big(\frac 23 L_{-2}L_{-2} + \frac{c-2}{16}L_{-4}\Big)\vac 
    \\
    \ket{\,[W^3]\,} &= \Big( 8 W_{-3}L_{-2} + \frac{7c-46}{8} W_{-5} \Big) \vac
    \label{eq:W2B5}
    \\
    \ket{\,[W^4]\,} &= 
    \Big(
    48 W_{-3}W_{-3}  + \frac{168}{9}L_{-2}L_{-2}L_{-2} 
    + \frac{57c + 206}{18} L_{-3}L_{-3} + \frac{2}{9} L_{-4}L_{-2} \nonumber\\&\qquad
    + \frac{513 c^2 + 3132 c -1148 }{288} L_{-6} \Big)\vac
\end{align}
The first three of these are equivalent (up to total derivatives) to integrable hierarchy currents,
\begin{align}
    \ket{\,[W]\,} &\sim \ket{J^B_3}
    \;,\;\;\\
    \ket{\,[W^2]\,} &\sim \frac 23 \,\ket{J_4^{\text{KdV}}}
    \label{eq:w2kdv4}
    \;,\;\;\\
    \ket{\,[W^3]\,} &\sim 8 \,\ket{J^B_5}
    \;,
    \end{align}
but the fourth is not.

The number of inequivalent bracketings in a commutative but non-associative product is given by the "Wedderburn-Etherington numbers", but so far we have not found an explicit expression for the coefficient in front of each inequivalent term.

These bracketings can also be neatly encoded in full binary (or Otter \cite{Otter}) trees, where each internal node corresponds to a bracketing, and each end leaf corresponds to a $W$, eg

\tikzset{
        blank/.style={draw=none},
         edge from parent/.style=
         {draw,edge from parent path={(\tikzparentnode) -- (\tikzchildnode)}},
         level distance=1.cm}

\[
(WW\lb  = \!\!\!\!
\raisebox{-12mm}{ 
    \begin{tikzpicture}
    \node at (-.4,0.0) {$W$};
    \node at (.4,0.0) {$W$};
    \fill (0,1) circle (2.4pt);
    \draw (0,1) -- (-.4,0.2);
    \draw (0,1) -- (.4,0.2);
\end{tikzpicture}}
,\;\,
(W(WW\lb\lb = \!\!\!\! 
   \raisebox{-2.cm}{
    \begin{tikzpicture}
        \node at (-.7,0.0) {$W$};
    \node at (.3,-.8) {$W$};
    \node at (1.1,-.8) {$W$};
    \fill (0,1) circle (2.4pt);
    \draw (0,1) -- (-.7,0.2);
    \draw (0,1) -- (.7,0.2);
\fill (.7,.2) circle (2.4pt);
    \draw (.7,.2) -- (1.1,-0.6);
    \draw (.7,.2) -- (.3,-0.6);
\end{tikzpicture}}
\!\!\!\!,\;\,
((WW\lb (WW)\lb\lb = \!\!\!\!\!\!\!
   \raisebox{-2.cm}{
    \begin{tikzpicture}
    \node at (.3,-.8) {$W$};
    \node at (1.1,-.8) {$W$};
    \node at (-.3,-.8) {$W$};
    \node at (-1.1,-.8) {$W$};
    \fill (0,1) circle (2.4pt);
    \draw (0,1) -- (-.7,0.2);
    \draw (0,1) -- (.7,0.2);
\fill (-.7,.2) circle (2.4pt);
    \draw (-.7,.2) -- (-1.1,-0.6);
    \draw (-.7,.2) -- (-.3,-0.6);
\fill (.7,.2) circle (2.4pt);
    \draw (.7,.2) -- (1.1,-0.6);
    \draw (.7,.2) -- (.3,-0.6);
\end{tikzpicture}}
\;.
\]
This makes the recursion relation \eqref{eq:conj1b} clear - the trees in $[W^n]$ can be split into those that have $m$ nodes $W$ on the left and $(n-m)$ on the right. There is a unique initial stage of generating these tress which is to add two nodes to the initial node, leaving $(n-2)$ nodes to be added. Of these, $m-1$ have to be added to the left mode and $n-m-1$ to the right node. There are  $\binom{n-2}{m-1}$ ways of choosing whether the next node is added to the left or right side, so that this term comes with that factor, leading to the recursion relation \eqref{eq:conj1b}.
\begin{align}
    [W^n] = 
    \raisebox{-1.4cm}{
    \begin{tikzpicture}
    \draw (0,0) circle (5mm);
    \node at (0,0) {$\scriptstyle{n}$};
    \fill (0,1) circle (2.4pt);
    \draw (0,1) -- (-.46,0.2);
    \draw (0,1) -- (.46,0.2);
\end{tikzpicture}}
=    \;
    \sum_{m=1}^{n-1}
    \binom {n-2}{m-1}
    \raisebox{-2.4cm}{
    \begin{tikzpicture}
    \fill (0,1) circle (2.4pt);
    \draw (0,1) -- (-.7,0.2);
    \draw (0,1) -- (.7,0.2);
    \fill (-.7,.2) circle (2.4pt);
    \draw (-.7,.2) -- (-1.16,-0.6);
    \draw (-.7,.2) -- (-.24,-0.6);
    \draw (-.7,-0.8) circle (5mm);
    \node at (-.7,-0.8) {$\scriptstyle{m}$};
\fill (.7,.2) circle (2.4pt);
    \draw (.7,.2) -- (1.16,-0.6);
    \draw (.7,.2) -- (.24,-0.6);
    \draw (.7,-0.8) circle (5mm);
    \node at (.7,-0.8) {${\scriptstyle{n{-}m}}$};
\end{tikzpicture}}
=  \;
    \sum_{m=1}^{n-1}
    \binom {n-2}{m-1}([W^m][W^{n-m}])
\end{align}
The method of generating the $n-th$ tree by adding a pair of nodes in every possible way leads to the following alternative recursive definition, 
\begin{align}
    [W^{k+1}] = 
    (WW) \frac{\partial}{\partial W}[W^k]
    \;,
\end{align}
which appears more suited to a quantum field theory interpretation since it implies
\begin{align}
\frac{\partial \CW}{\partial\alpha}
= \frac{\tau^2}{4\pi i} (WW) \frac{\partial \CW}{\partial W}
\;.
\end{align}

\section{Modular and quasi-modular forms}
\label{app:MF}

In this appendix we will list the relevant facts about modular forms that appear in this paper. Proofs of the following statements can be found in \cite{Zagier2008} and most of the notation will be the same. 

The modular group will be denoted by
\be
\Gamma_1 = \text{SL}(2,\mathbb{Z})/ \{\pm I\}\;,
\ee
Consider a matrix 
\be
    \begin{pmatrix}  a&b\\c&d\end{pmatrix} \in \Gamma_1 \;.
\ee 
If a holomorphic function $f(\tau)$, defined in the upper half plane, has the following transformation property 
\be\label{eq:MF def}
f\left(\frac{a\tau+b}{c\tau+d}\right)=(c\tau+d)^kf(\tau)\;,
\ee
then we say that the function is a holomorphic modular form of weight
$k$ on $\Gamma_1$. We will denote the space of modular forms of weight
$k$ on $\Gamma_1$ by $M_k(\Gamma_1)$. 

The group $\Gamma_1$ is finitely generated by the matrices
\be
    \pm \begin{pmatrix}1&1\\0&1\end{pmatrix} \;,\quad \pm\begin{pmatrix} 0&1\\-1&0 \end{pmatrix}\;,
\ee 
hence we only need to check that a function transforms as a modular form under
\be
    T:\tau\mapsto\tau+1 \;,\quad S:\tau\mapsto\frac{-1}{\tau}\;,
\ee
to verify it is an element of $M_k(\Gamma_1)$. 

An important fact about the space $M_k(\Gamma_1)$ is that it is finite dimensional. The space $M_{2k}(\Gamma_1)$ is generated by the Eisenstein series $E_4$ and $E_6$, which we now define. 

The Eisenstein series $E_{2k}(\tau)$ are elements of $M_{2k}(\Gamma_1)$ for $k=2,3,\dots$ and they are defined by
\be
E_{2k}(\tau)=1+\frac{2}{\zeta(1-2k)}\sum_{n=0}^\infty\frac{n^{2k-1}q^n}{1-q^n}\;,\;\;\;q=e^{2\pi i\tau}\;.
\ee
The Eisenstein series also satisfy differential identities,
\begin{align}
    E_2' \equiv q\frac{d}{dq} E_2 = 
    \frac{E_2^2 - E_4}{12}
    \;,\;\;
    E_4' \equiv q\frac{d}{dq} E_4 = 
    \frac{E_2E_4 - E_6}{3}
    \;,\;\;
    E_6' \equiv q\frac{d}{dq} E_6 = 
    \frac{E_2E_6 - E_4}{2}
    \;.   
    \label{eq:Enderivs}
\end{align}
For $k=1$ the Eisenstein series $E_2(\tau)$ is quasi-modular which means that under a modular transform we have the transformation property
\be
E_{2}\left(\frac{a\tau+b}{c\tau+d}\right)=(c\tau+d)^2E_2(\tau)-\frac{6i}{\pi}c(c\tau+d)\;,
\ee
and in particular
\be
E_2(-\tfrac 1\t) = \t^2 E_2 - \frac {6i\t}{\pi}
\;.
\ee

We also encounter quasi-modular forms. For our purpose we will define the space of quasi-modular forms of weight $k$ and depth $p$, denoted by $\tilde  M_k^{(\leq p)}(\Gamma_1)$, to be
\be\label{eq:QMF def}
    \tilde M_k^{(\leq p)}(\Gamma_1) = \bigoplus_{r=0}^pM_{k-2r}(\Gamma_1)\cdot E_2^r \;,
\ee 
where the coefficient of $E_2^p$ is non-zero.

Finally we define the Serre derivative. The Serre derivative acting on a modular form $f(\tau)$ of weight $k$ is defined to be
\be
    \SD k f(\tau)=\frac{1}{2\pi i}\frac{d}{d\tau}f(\tau)-\frac{k}{12}E_2(\tau)f(\tau)\;.
\ee
By using the transformation of $\frac{d}{d\tau}$ under a modular transform we can see that $\SD{k}f(\tau)$ is a modular form of weight $k+2$.

We can also use \eqref{eq:Enderivs} to  find the transformation properties of $E_2', E_4'$ and $E_6'$ easily:
\begin{align} 
    E_2'(-\tfrac 1\t)
    &= \t^4E_2'(\t)  - \frac{i}{\pi}\t^3 E_2(\t) -\frac{3}{\pi^2}\t^2
    \;,
    \label{eq:E2'trans}
    \\
    E_4'(-\tfrac 1\tau)
    &= 
    \tau^6 E_4'(\t)  - \frac{2i}{\pi}\tau^5 E_4(\t)
    \;,
    \;\;
        \label{eq:E4'trans}
\\    
    E_6'(-\tfrac 1\tau)
    &= 
    \tau^8 E_6'(\t)  - \frac{3i}{\pi}\tau^7 E_6(\t)\;.
        \label{eq:E6'trans}
\end{align}
Finally it is helpful to note 
the definition and derivative of $\eta$, 
\be\label{eq:eta derivative}
\eta(q) = q^{-1/24}\prod_{n=1}^\infty(1-q^n)
\;,\;\;
q \frac{d}{d q} \eta = \frac{E_2}{24} \eta
\;.
\ee

\section{Zhu Recursion Relation}
\label{app:Zhu}
In this appendix, we will briefly review the Zhu recursion relation. A detailed review of Zhu recursion relation can be found in \cite{Gaberdiel:2012yb} and various examples can be found in \cite{Ashok:2024zmw}.

We define the $n$-point correlator on a torus using the conventions of \cite{Gaberdiel:2012yb} as 
\begin{equation}
        F_r((a^1,z_1) \cdots (a^n,z_n) ; \tau) = \text{Tr}_r\,[ V(a^1,z_1) \cdots V(a^n,z_n) q^{L_0-\frac{c}{24}}] \prod_{i=1}^n z^{\wgt(a_i)} \;,
\end{equation}
\noindent where the trace is over any representation $r$ of the vertex algebra, $z_i=e^{2\pi i u_i}$ where $u_i$ is the coordinate on the torus\footnote{Note the parametrisation of the torus here is different to that in appendix \ref{app:Gaberdiel}, to conform to the conventions of \cite{Gaberdiel:2012yb}}, $a^i$ is a state, $\wgt(a_i)$ is the conformal dimension of the state, and $V(a^i,z)$ is the vertex operator (field) corresponding to the state $a^i$. Eg, if $a^i = W_{-3}\vac$, then $\wgt(a^i)=3$ and $V(a^i,z) = W(z)$.

For a given $n$-point function on the torus, Zhu recursion relation allows us to compute it recursively in terms of the traces of the zero modes of the fields present in the correlator \cite{zhu1990vertex}.
The basic idea that underlies this formula is the following fact: commuting the $m^{\text{th}}$ mode of a local field across other local fields and then over $q^{L_0}$ in a trace gives us the original correlator with a factor of $q^m$, modulo terms proportional to $(n-1)$-point correlators. The commutators involved follow from projecting the required mode from the OPE of the local fields:
\be
[a_k, V (b,z)] = \sum_{n\geq 0} \binom{\wgt(a)+1-k}{n} V(a_{n-\wgt(a)+1}b,z)z^{\wgt(a)+1-k-n}~,
\ee
where the plane modes $a_n$ arise from a formal series expansion of the operator associated with a state $a$ in terms of the plane coordinate $z$
 \be
V(a,z)=\sum_{n\in\mathbb{Z}}\frac{a_{n}}{z^{n+\wgt(a)}}~,
\ee
Since we are working on the torus, where vertex operators naturally take the form \\$V(q^{L_0}v, e^{2\pi i u})$, it is natural that terms arising from the commutator can be reorganized into terms of the kind:
\begin{align}
    \langle V(q_u^{L_0}a_1, q_u)V(q_v^{L_0}a_2, q_v); \tau \rangle &= \langle V(V(q_u^{L_0}a_1, q_u-q_v)q_v^{L_0}a_2, q_v); \tau \rangle \\ &= \langle V(q_v^{L_0}V(q_{u-v}^{L_0}a_1, q_{u-v}-1)a_2, q_v); \tau \rangle \;,
\end{align}
where $q_x := e^{2\pi i x}$. The modes of the operator $V(q_{u-v}^{L_0}a_1, q_{u-v}-1)$ are the square modes that will appear in the recursion relations.
\be
    V(q_x^{L_0}v,q_x-1) = \sum_{n\in \Zint} v[n] \, x^{-n-1} \;.
\ee
Explicitly, the square bracket modes can be written in terms of plane modes as:
 \be
 \label{eq:squaremodes1}
 a[n]=\frac{1}{(2\pi i)^{n+1}}\sum_{j\ge n+1-h_a}c(h_{a},j+h_{a}-1,n)a_{j}~,
 \ee
where the expansion coefficients are given by
  \be
  \label{eq:coeff}
  (\log(1+z))^{s}(1+z)^{h-1}=\sum_{j\geq s}c(h,j,s)z^{j}~.
  \ee
This formula will be used, for instance, to compute the square modes of $W(z)$, whose vertex operator representation is given by  $V(W_{-3}\vac, z)$.
The Jacobi identity for square modes  (see equation $(4.2.4)$ of \cite{Zhu:1996gaq}) is useful for computing the square modes of composite objects:
\be
\label{eq:zhusquare}
(b[n]a)[m]=\sum_{i}\binom{n}{i}\Big((-1)^ib[n-i]a[m+i]-(-1)^{n+i}a[n+m-i]b[i]\Big)~.
\ee
The Zhu recursion formula is given by \cite{Gaberdiel:2012yb}:
\be
\begin{aligned}
\label{eq:zhu}
&F(b_{0}^{l};(a^{1},z_1),...,(a^{n},z_{n});\tau) = F( b_{0}^{l}a_{0}^{1};(a^{2},z_2),...,(a^{n},z_{n});\tau )  \\&  + \, \sum_{i=0}^{l}\sum_{j=2}^{n}\sum_{m=0}^{\infty}\binom l ig^{i}_{m+1}\left(\frac{z_{j}}{z_1}, q\right) F( b_{0}^{l-i};(a^{2},z_2),...,(d^{i}[m]a^{j},z_{j}),...,(a^{n},z_{n});\tau ) \;,
\end{aligned}
\ee
where $d^i[m]$ are defined through :
\be
    d^i[m] = (-1)^i (b[0])^i a^1 \;,
\ee
and $g^j_k(x,q)$ are :
\be
g^{j}_{k}(x,q)=\frac{(2\pi i)^{k}}{(k-1)!}\sum_{n\neq 0}n^{k-j-1}x^{n}\partial^{j}\frac{1}{(1-q^n)}=(2\pi i)^{2j}\frac{(k-j-1)!}{(k-1)!}\partial^j\mathcal{P}_{k-i}(x,q)~.
\ee
where $\partial^j:=(q\partial_q)^j$. $\CP_1(x,q)$ is related to the Weierstrass $\wp$ function  through \cite{zhu1990vertex}:
\be
\begin{aligned}
\label{eq:weiersPrho}
\mathcal{P}_1(e^{2\pi i u},q) = -\wp(u,\tau) + \frac{\pi^2}{3} u E_2(\tau) - i\pi  \;,
\end{aligned}
\ee
and the $\mathcal{P}_k$ for $k>1$ can be obtained through the relations:
\begin{align}
    \p_u \mathcal{P}_k(e^{2\pi i u},q) &= k \mathcal{P}_{k+1}(x,q) \;.\\
\end{align}
Although \eqref{eq:zhu} only allows for the insertion of zero modes of a single kind, a generalization that allows for zero modes of more than one kind into the recursion relation appeared recently in \cite{Addabbo:2024ljn}.

\section{List of integrals}
\label{app:c}

In this section, we list all the integral involving Weierstrass functions which are relevant for the computations in section \ref{sec:4}.  
\subsection{Integrals involving one  Weierstrass function}
\label{sec:C1}
\vspace{0.5 cm}

      \begin{equation}\int_{1}^{q}\frac{dz_i}{z_i}\mathcal{P}_{1}\Big(\frac{z_k}{z_i}\Big)=(2\pi i)(i\pi-2\pi i\tau+\log(z_k))\end{equation}
   
   \begin{equation}
        \int_{1}^{q}\frac{dz_k}{z_k}\mathcal{P}_{1}\Big(\frac{z_k}{z_i}\Big)=-(2\pi i)(i\pi+\log(z_k))
    \end{equation}
    \begin{equation}
        \int_{1}^{q}\frac{dz_i}{z_i}\mathcal{P}_{2}\Big(\frac{z_k}{z_i}\Big)=(2\pi i)^2
          \end{equation}
        \begin{equation}
        \int_{1}^{q}\frac{dz_i}{z_i}\mathcal{P}_{m}\Big(\frac{z_k}{z_i}\Big)=0~,~~m>2
    \end{equation}
  \begin{equation}
        \int_{1}^{q}\frac{dz_k}{z_k}\mathcal{P}_{2}\Big(\frac{z_k}{z_i}\Big)=(2\pi i)^2
          \end{equation}
           \begin{equation}
        \int_{1}^{q}\frac{dz_i}{z_i}\log(z_i)\mathcal{P}_{m+1}\Big(\frac{z_i}{z_k}\Big)=(2\pi i)^2\frac{\tau}{m}\mathcal{P}_{m}\Big(\frac{1}{z_k}\Big)-(2\pi i)^3\frac{1}{m}\delta_{m,2}~~~~m\geq2
          \end{equation}
           \begin{equation}
        \int_{1}^{q}\frac{dz_k}{z_k}\partial_\tau\mathcal{P}_{m}\Big(\frac{z_i}{z_k}\Big)=-(2\pi i)\mathcal{P}_{m}\Big(z_i\Big)~~~~m>1
          \end{equation}
            \begin{equation}
        \int_{1}^{q}\frac{dz_k}{z_k}\mathcal{P}_{2}\Big(z_k\Big)=(2 \pi i)^2
          \end{equation}
         \begin{equation}
        \int_{1}^{q}\frac{dz_k}{z_k}\mathcal{P}_{1}\Big(z_k\Big)=0
          \end{equation}
           \begin{equation}
        \int_{1}^{q}\frac{dz_k}{z_k}\partial^{2}_{q}\mathcal{P}_{1}\Big(z_k\Big)=0~
          \end{equation}
             \begin{equation}
        \int_{1}^{q}\frac{dz_k}{z_k}\partial_{q}\mathcal{P}_{1}\Big(z_k\Big)=0
          \end{equation}
         \begin{equation}
          \begin{aligned}
                \int_{1}^{q}\frac{dz_k}{z_k}\int_{1}^{q}\frac{dz_j}{z_j}\partial^2_{q}\mathcal{P}_{m}\Big(\frac{z_k}{z_j}\Big)=&\frac{2(2\pi i )^m}{(m-1)!}\zeta(1-m)E_m(\tau)~~~\text{when}~m~\text{is even.}\\
                &=0~~~\text{when}~m~\text{is odd.}
          \end{aligned}
          \end{equation}

\subsection{Integrals involving two Weierstrass functions}
\label{sec:C2}
Besides those given in \ref{sec:C1}, some non-trivial integrals that we encounter are listed below:
 \begin{equation}
    \begin{aligned}
    \int_{1}^{q}\frac{dz_i}{z_i} \int_{1}^{q}\frac{dz_j}{z_j} \int_{1}^{q}\frac{dz_k}{z_k}\mathcal{P}_{1}\Big(\frac{z_i}{z_j}\Big)\partial_q\mathcal{P}_{m-1}\Big(\frac{z_k}{z_i}\Big)&= \frac{(2\pi i)^{m+2}}{(m-2)!} \tau^2 \zeta(3-m)E_{m-2} \\&+\frac{2(2\pi i)^{m+1}}{(m-2)!}  \zeta(4-m),~~~\text{when}~m~\text{is even.}\\
    \end{aligned}
    \end{equation}
     \begin{equation}
    \begin{aligned}
    \int_{1}^{q}\frac{dz_i}{z_i} \int_{1}^{q}\frac{dz_j}{z_j} \int_{1}^{q}\frac{dz_k}{z_k}\mathcal{P}_{1}\Big(\frac{z_k}{z_j}\Big)\partial_q\mathcal{P}_{m-1}\Big(\frac{z_k}{z_i}\Big)&= -\frac{(2\pi i)^{m+2}}{(m-2)!}  \tau^2 \zeta(3-m)E_{m-2} \\&-\frac{2(2\pi i)^{m+1}}{(m-2)!}  \zeta(4-m),~~~\text{when}~m~\text{is even.}\\
    \end{aligned}
    \end{equation}

\subsection{Integrals involving three Weierstrass functions}
 \begin{equation}
    \begin{aligned}
   &\int_{1}^{q}\frac{dz_l}{z_l} \int_{1}^{q}\frac{dz_i}{z_i} \int_{1}^{q}\frac{dz_j}{z_j} \int_{1}^{q}\frac{dz_k}{z_k}\mathcal{P}_{1}\Big(\frac{z_i}{z_j}\Big)\mathcal{P}_{1}\Big(\frac{z_k}{z_l}\Big)\mathcal{P}_{m}\Big(\frac{z_k}{z_i}\Big)\\&~~~~~~=-(2\pi i\tau)\frac{(2\pi i)^{m+3}}{(m-1)!}  \tau ~\zeta(3-m)E_{m-2}-(2\pi i\tau)\frac{2(2\pi i)^{m+2}}{(m-1)!}  \zeta(4-m)~~~\text{when}~m~\text{is even.}\\
    \end{aligned}
    \end{equation}

\section{Modular transformations from free field representations}
\label{app:Verma}

It was shown in \cite{Iles:2013jha} how modular transformations of expressions of the form
\begin{align}
 p(\htilde,w)\,\chi(\hat q)    
\end{align}
could be found using a free field representation of the $\CW_3$ algebra. In that paper it was only applied to 
$ w\,\chi$ but it can be used for any polynomial, as we explain now.

We start from the fact that 
Quantum conformal Toda field theories can be associated to any
finite-dimensional Lie algebra $g$ and are theories constructed from
$r = \mathrm{rank}(g)$ bosonic scalar fields which depend on a coupling
constant and 
which have $\CW$-algebra symmetries \cite{Mansfield:1990du}. 
We can restrict attention to the modular invariant set of $\CW_3$ representations labelled by two integers $(a_1,a_2)$ for which 
\be
  \htilde = h - \frac{c-2}{24} = \frac{a_1^2 + a_1a_2 + a_2^2}3 + \frac{c}{24}
\;,\;\;\;\;
  w = \frac 1{27}
     (a_1{-}a_2)(2a_1 {+} a_2)(a_1 {+} 2 a_2)
\;.
\label{eq:wts}
\ee
For $c>98$ and $a_1,a_2$ real, these correspond to Verma modules.
If we change variables from $(a_1,a_2)$ to $(p_1,p_2)$, 
\be
 {\bf p} = (p_1,p_2)
 = \left( \frac{a_1}{\sqrt 2}, \frac{a_1}{\sqrt 6} + \sqrt{\frac{2}{3}}a_2
   \right)
\;,
\ee
then 
\be
\htilde(a_1,a_2) = \frac 12 {\bf p}^2
\;,\;\;
w(a_1,a_2) = \frac{1}{3\sqrt 6} p_2(p_1^2 - p_2^2)
\;,
\ee
and the character of the representation is
\be
\chi_{h(a_1,a_2)} = \frac{q^{\htilde}}{\eta(q)^2} = \frac{ q^{\frac 12 {\bf p}^2}}{\eta(q)^2}
\ee
Furthermore, if we write
\be{\bf d}\cdot{\bf p}
= d_1 p_1 + d_2 p_2\;,
\ee
then 
\begin{align}
  \htilde({\bf p})\,\exp({\bf d} \cdot {\bf p})
&= \frac 1{2}(p_1^2 + p_2^2)\exp({\bf d}\cdot{\bf p})
\nonumber \\
&= \frac{1}{2}
\left( \frac{\partial^2}{\partial d_1^2} +  \frac{\partial^2}{\partial d_2^2}\right)\exp({\bf d} \cdot{\bf p})
\nonumber \\
&= {\CD_h} \exp({\bf d} \cdot {\bf p})
\\
w({\bf p}) \exp({\bf d} \cdot {\bf p})
&= \frac 1{3\sqrt 6}(p_2(p_1^2-p_2^2)) \exp({\bf d}\cdot{\bf p})
\nonumber \\&= \frac{1}{3\sqrt 6}
\left( 3 \frac{\partial^2}{\partial d_1^2} -  \frac{\partial^2}{\partial d_2^2} 
\right) \frac{\partial}{\partial d_2}\exp({\bf d} \cdot{\bf p})
\nonumber \\
&= {\CD_w} \exp({\bf d} \cdot {\bf p})
\;,
\end{align}
and so
\be
\htilde({\bf p})^m\,w({\bf p})^n
= (\CD_h)^m (\CD_w)^n \exp({\bf d}\cdot{\bf p}) \Big|_{{\bf d} = 0}
\;.
\label{eq:hwpolys}
\ee
Since the modular transform of the characters is given by the Fourier transform,
\be
  \chi_{\bf p}(\hat q)
= 
  \,\iint  e^{-2\pi i {\bf p}\cdot{\bf p'}}
   \chi_{\bf p'}(q) \, \rd^2p'
\;,
\label{eq:ppp}
\ee
we can now calculate the modular transform of $h({\bf p})^m\,w({\bf p})^n\,\chi_{\bf p}$ by expressing $h(\bf p)$ and $w(\bf p)$ in terms of differential operators on $\exp(\bf d \cdot \bf p)$ and the shifted Fourier transform
\be
\exp( - \pi i {\bf p}^2/\tau + {\bf d}\cdot{\bf p})
= 
(-i\tau)
\,
\iint \exp( \pi i \tau {\bf p'}^2 - 2 \pi i {\bf p}\cdot{\bf p'})
      \exp( -\frac{i\tau}{4\pi} {\bf d}^2 + {\bf d}\cdot{\bf p'}\tau )
\, \rd^2 p'
\;.
\label{eq:ftshifted}
\ee
This gives
\begin{align}
\htilde({\bf p})^m w({\bf p})^n\,\chi_{\bf p}(\hat q)
&= 
\left.
(\CD_h)^m\,(\CD_w)^n\,
\frac{\exp(-\pi i{\bf p}^2/\t + {\bf d}\cdot{\bf p})}{\eta(\hat q)^2}
\right|_{{\bf d}=0}
 \nonumber\\
 &=
 \iint e^{-2\pi i {\bf p}\cdot{\bf p}'}\,
 \chi_{\bf p'}(q)
 \,\left.
(\CD_h)^m\,(\CD_w)^n\,
 \exp( - \frac{i\t}{4\pi}{\bf d}^2 + {\bf d}\cdot{\bf p}')
 \,\rd^2 p'\;
 \right|_{\bf d=0} \;,
\end{align}
or simply
\begin{align}
   \htilde({\bf p})^m w({\bf p})^n\,\chi_{\bf p}(\hat q) 
   \;\to \;\left.
(\CD_h)^m\,(\CD_w)^n\,
 \exp( - \frac{i\t}{4\pi}{\bf d}^2 + {\bf d}\cdot{\bf p}')
 \,\rd^2 p'\;
 \right|_{\bf d=0}
 \,
   \chi_{\bf p'}(q)
 \;.
\end{align}
Applying this in a few simple cases we get
\begin{align}
    \chi(\hat q) & \to \chi(q)
    \;,\;\;
    \nonumber\\
    \tilde h \chi(\hat q) & \to (\tau^2 \tilde h - \frac i{2\pi}\t ) \chi(q)
    \;,\;\;
    \nonumber\\
    w \chi(\hat q) & \to \tau^3 w \chi(q)
        \;,\;\;
    \nonumber\\
    \tilde h^2 \chi(\hat q) & \to 
    (\tau^4 \tilde h^2 - \frac{2i}{\pi}\tau^3\tilde h - \frac{1}{2\pi^2}\tau^2 ) \chi(q)
        \;,\;\;
    \nonumber\\
    \tilde h w \chi(\hat q) & \to 
    (\tau^5 \tilde h w  - \frac{2i}{\pi}\tau^4 w ) \chi(q)
        \;,\;\;
    \nonumber\\
    w^2 \chi(\hat q) & \to 
    (\tau^6 w^2 - \frac{i}{3\pi}\tau^5\tilde h^2 - \frac{1}{3\pi^2}\tau^4\tilde h + \frac{i}{18 \pi^3}\tau^3 ) \chi(q)
        \;,\;\;
    \nonumber\\
    \tilde h^3  \chi(\hat q) & \to 
    (\tau^6 \tilde h^3 
    - \frac{9i}{2\pi}\tau^5 \tilde h^2
    -\frac{9}{2\pi^2}\tau^4\tilde h + \frac{3i}{4\pi^3}\tau^3 ) \chi(q)
        \;.
\label{eq:transformexamples}        
\end{align}
It is easy to check that these square to the charge conjugation operation $\{\tilde h \to \tilde h, w \to -w\}$, as expected.

\section{Relation to the work of Dijkgraaf}
\label{app:Dijkgraaf}

In \cite{Downing:2021mfw}, the authors were able to use the work of Dijkgraaf \cite{Dijkgraaf:1996iy} to find the asymptotic form of the relation 
\eqref{eq:conj1} and correspondingly the transforms of the individual traces \eqref{eq:TrW0n}. Unfortunately this method is not open to us as the currents do not satisfy the pre-algebra condition that the first order pole terms are total derivatives. 
Taking the first two currents to be $W(z)$ and $[W^2](z) = (2/3)\Lambda(z)$ 
we find their OPE is 
\begin{align}
    W(z)\,\,[W^2](w)
    &= \frac{32}{5}\Big(
    \frac{W(w)}{(z-w)^4}
    + \frac 13 \frac{W'(w)}{(z-w)^3}
    + \frac{1}{7}\frac{W''(w)}{(z-w)^2}
    + \frac{1}{14}\frac{W'''(w)}{(z-w)}\Big)
    \nonumber\\
    &+
    \frac{1}{9}\Big(
    \frac{K(w)}{(z-w)^2} + \frac 25 \frac{K'(w)}{(z-w)}\Big)
    \;-\;
    \frac{8}{5}\frac{J(w)}{(z-w)} \; + O(1)\;, 
    \label{eq:WW2OPE}
\end{align}
where the fields $K$ and $J$ are quasi primary fields corresponding to the states (in the notation of \cite{Kausch:1990bn},
\begin{align}
    \ket K &= \Lambda^{(3)}_{-2}\,W_{-3}\vac
    \nonumber\\
    &= 
    \Big(
    L_{-2}W_{-3}
    -\frac{3}{14}W_{-5}
    \Big)\vac
    \;,\\
    \ket J &= 
    \Lambda^{(3)}_{-3}\,W_{-3}\vac
    \nonumber\\
    &=
    \Big(
    L_{-3}W_{-3} 
    -\frac 23 L_{-2}W_{-4}
    + \frac 1{2}W_{-6}
    \Big)\vac
    \;.
    \label{eq:Jdef}
\end{align}
As can be seen, the first order pole in this OPE has a contribution from the quasi-primary field $J$ and so is not a total derivative. As a consequence, the pre-algebra structure used in \cite{Dijkgraaf:1996iy} does not exist and we cannot use his results.

\section{The map \texorpdfstring{$\bf{\hat{}}$}{\bf{\hat{}}} from the cylinder to the plane}
 
\label{app:Gaberdiel}

Throughout this paper we use the hat notation to indicate a specific map from fields on the cylinder (of circumference $2\pi$) to the plane, $A(u)\mapsto \hat A(z)$ and the zero-mode notation to indicate a specific map in these fields $A(z) \to \hat A_0$. 

In \cite{Gaberdiel:1994fs}, Gaberdiel gives a formula for the transformation of fields under a general conformal map. For the map from the cylinder $0\leq \text{Im}\,u<2\pi$ to the plane $z$, $z = \exp (u)$, a quasi-primary field $\psi(z)$ of weight $h$ is mapped to a sum of fields of $\psi_n$ weights $h_n = h-n$, and $\psi_0 = \psi$,
\be
\psi(u) \mapsto \sum_{n=0}^h z^{h -n} \, \psi_n(z)
\;,
\ee
for example
\begin{align}
    T(u) \mapsto z^2 T(z) - \frac{c}{24}
    \;,\;\;
    \Lambda(u) \mapsto
    z^4 \Lambda(z) 
    - \frac{5c+22}{60}z^2T(z) + \frac{c(5c+22)}{2880}
    \;.
\end{align}
The result is that the integrals of the fields on the cylinder along the line $\text{Re}(u)=0$, map to zero modes of the fields on the plane. With $u=x+iy$,
\begin{align}
    \int_{y=0}^{2\pi} T(u) \frac{\rd y}{2\pi}
    &= \int T(u) \frac{\rd u}{2\pi i}
    \oint \frac{\rd z}{2\pi i z}
    =
    (z^2 T(z) - c/24)
    \nonumber\\
    &= L_0 - c/24
    \;,
    \\
    \int_{y=0}^{2\pi} \Lambda(u) \frac{\rd y}{2\pi}
    &= 
    \oint \frac{\rd z}{2\pi i z}
    (z^4 \Lambda(z) - \frac{5c{+}22}{60}z^2 T(z) + \frac{c(5c{+}22)}{2880})
    \nonumber\\
    &= \Lambda_0 - \frac{5c{+}22}{60}L_0 + \frac{c(5c{+}22)}{2880}
    \;.   
\end{align}
In order to avoid repeated powers of $z$ in all the expressions obtained from maps to the plane, we use $\hat{A}$ to indicate the map of the field $A$ with the pre-factors suppressed, and the same for the state $\ket{\hat A}$, and the zero mode $\hat A_0$ to indicate the integral of the field $A$ on the cylinder, after the map to the plane, eg
\begin{align}
\hat\Lambda(z) &=
\Lambda(z) - \frac{5c{+}22}{60}T(z) + \frac{c(5c{+}22)}{2880}\;,\\
\ket{\hat \Lambda} &=
\Lambda_{-4}\vac - \frac{5c{+}22}{60}L_{-2}\vac + \frac{c(5c{+}22)}{2880}\vac\;,\\
\hat\Lambda_0
&=\Lambda_0 - \frac{5c{+}22}{60}L_0 + \frac{c(5c{+}22)}{2880}
    \;.   
\end{align}
These are just for notational convenience, with the added advantage that $\hat A(z)$ corresponds to the state $\ket{\hat A}$ by the field-state correspondence, ie in Vertex operator language,
\be
V(\,\ket{\hat A},z,s) = \hat A(z)
\;,
\ee
and $\hat A_0$ is the zero more of the field $\hat A(z)$.


\newpage

\bibliographystyle{JHEP}
\bibliography{bib}
\end{document}